\documentclass[twocolumn,showpacs,amsmath,amssymb,prc,aps,10pt]{revtex4}

\usepackage{graphicx}
\usepackage{float}
\usepackage{dcolumn}
\usepackage{bm}
\usepackage{multirow}
\usepackage{amssymb}
\usepackage{amsmath}
\usepackage[usenames,dvipsnames]{xcolor}

\newcommand{\Omegabar}{\ensuremath{\overline{\Omega}}}
\newcommand{\Xibar}{\ensuremath{\overline{\Xi}}}
\newcommand{\MeV}{\ensuremath{\,{\rm MeV}}}
\newcommand{\GeV}{\ensuremath{\,{\rm GeV}}}
\newcommand{\TeV}{\ensuremath{\,{\rm TeV}}}
\newcommand{\req}[1]{Eq.\,\ref{#1}}
\newcommand{\rf}[1]{figure~\ref{#1}}

\begin{document}
\topmargin -1.0cm

\preprint{CERN-PH-TH/2012-262}

\title{Hadron production and QGP Hadronization in Pb--Pb collisions at $\sqrt{s_{NN}}=2.76$ TeV}

\author{Michal Petr\' a\v n$^{1,3}$}
\author{Jean Letessier$^2$}
\author{Vojt{\v e}ch Petr{\' a}{\v c}ek$^3$}
\author{Johann Rafelski$^{1,4}$}%
\affiliation{%
$\quad$\\ $^1$Department of Physics, The University of Arizona, Tucson, Arizona 85721, USA 
}%
\affiliation{
$^2$Laboratoire de Physique Th{\' e}orique et Hautes Energies, Universit{\' e} Paris 6, Paris 75005, France
}%
\affiliation{
$^3$Czech Technical University in Prague, Faculty of Nuclear Sciences and Physical Engineering
}%
\affiliation{
$^4$Theory Division of Physics Department, CERN, CH-1211 Geneva 23, Switzerland
}%

\date{March 8, 2013}

\begin{abstract}
We show that all central rapidity hadron yields measured in Pb--Pb collisions at $\sqrt{s_{NN}}=2.76$ TeV are well described by the  chemical non-equilibrium statistical hadronization model ({\small SHM}), where the  chemically equilibrated {\small QGP} source  breaks up  directly into hadrons. {\small SHM}  parameters are obtained as a function of centrality of colliding ions, and we compare  CERN Large Hadron Collider ({\small LHC}) with  Brookhaven National Laboratory Relativistic Heavy Ion Collider ({\small RHIC}) results.   We predict yields of unobserved hadrons and address anti-matter production. The physical properties of the   quark--gluon plasma fireball particle source  show universality of hadronization conditions at  {\small LHC} and  {\small RHIC}.
\end{abstract}

\maketitle

\section{\label{sec:intro}Introduction and motivation}
\vskip-10.5cm
CERN-PH-TH/2012-262
\vskip-\baselineskip 
\vskip10.5cm
Our interest in the multi-particle production process in ultra-relativistic heavy ion collisions originates in the understanding that the transverse momentum integrated  rapidity distributions are insensitive to the very difficult to fully characterize  transverse evolution dynamics of the hot fireball source~\cite{Bozek:2012qs}.  A successful description of central rapidity particle yields in a single freeze-out model~\cite{Rybczynski:2012ed,Broniowski:2001we} will be used here to characterize the properties of the  hadronizing quark--gluon plasma ({\small QGP}) fireball. The  {\small QGP} breakup, as modeled within  the statistical hadronization model ({\small SHM}),  assumes equal reaction strength in all hadron particle production channels. Therefore, the phase space volume  determines  the hadron yields.  {\small SHM} has been described extensively before and we refer the reader to {\small SHARE} manuals~\cite{Torrieri:2004zz} for both, further theoretical details,  and numerical methods. Here, we apply {\small SHM} to study particle production in Pb--Pb collisions at $\sqrt{s_{NN}}=2.76\TeV$ ({\small LHC}2760), a new energy domain an order of magnitude higher than previously explored in Au--Au collisions at $\sqrt{s_{NN}}=200$ GeV ({\small RHIC}200). 

We begin by demonstrating that the chemical non-equilibrium {\small SHM} variant describes the experimental {\small LHC}-ion data with high accuracy. This finding disagrees with claims that {\small SHM}  alone does not describe the particle multiplicity data obtained in relativistic heavy ion collisions at {\small LHC}~\cite{Abelev:2012vx,Abelev:2012wca}. In the chemical non-equilibrium {\small SHM} approach, we allow quark pair yield parameter $\gamma_q$ for light quarks, a feature we presented as necessary model refinement for the past 15 years~\cite{Letessier:1998sz,Letessier:2000ay,Letessier:2002gp}.  We demonstrate the general model validity in our numerical approach by showing   correspondence  of  chemical equilibrium {\small SHM} results with other fits to the {\small LHC} data. This  demonstrates that several {\small SHM} programs, which had years time to mature and evolve, are compatible in their data tables of hadronic resonance mass spectra and decay patterns. However, only our extended {\small SHARE}-code includes advanced features, such as chemical non-equilibrium of all quark yields, differentiation of up and down quarks, evaluation of fireball physical properties, and the capability to constrain the fit by imposing  physical properties on the particle source.

To demonstrate that our chemical non-equilibrium {\small SHM}  works at {\small LHC}, we show in the left panel of \rf{fig:firstfit}  our fit to the 0--20\% centrality data, shown in the second column of table~\ref{tab:data}, recently presented and studied  by  the experiment {\small ALICE}~\cite{Abelev:2012vx,Abelev:2012wca}. Only in this one instance, we consider the relatively wide centrality trigger of 0--20\% to compare directly with the earlier analysis effort. As can be seen in the left panel of \rf{fig:firstfit}, our non-equilibrium {\small SHM} approach describes  these data with $\chi^2/\mathrm{ndf}=9.5/9\simeq1$. We see, in \rf{fig:firstfit}a insert, that the chemical equilibrium  {\small SHM} works poorly, $\chi^2/\mathrm{ndf} =64/11\simeq 6$, which is the same finding and conclusion as in~\cite{Abelev:2012vx,Abelev:2012wca}. 

While the equilibrium {\small SHM}  disagrees at {\small LHC} across many particle yields  the most discussed data point is  the p$/\pi = 0.046\pm0.003$ ratio~\cite{Abelev:2012wca}, a point we will study in more detail in subsection \ref{sec:pOverPi}.  Our work shows that the inability of the equilibrium {\small SHM} alone to fit the experimental value of p/$\pi$ ratio   does not mean that all variants of {\small SHM}  do not describe particle production in heavy ion collisions at {\small LHC}. One of the key findings of this work is that the chemical non-equilibrium {\small SHM} variant without any additional post-hadronization evolution provides an excellent description of all data. We will also argue that the present day hybrid models, that is models which combine SHM results with post-hadronization hadron yield evolution, need to address key features of the data such as quasi-constancy of the p/$\pi$ ratio as a function of centrality of the heavy ion collision and the abundance of multi-strange baryons.
 
\begin{figure*}[h!t]
\includegraphics[height=12.cm]{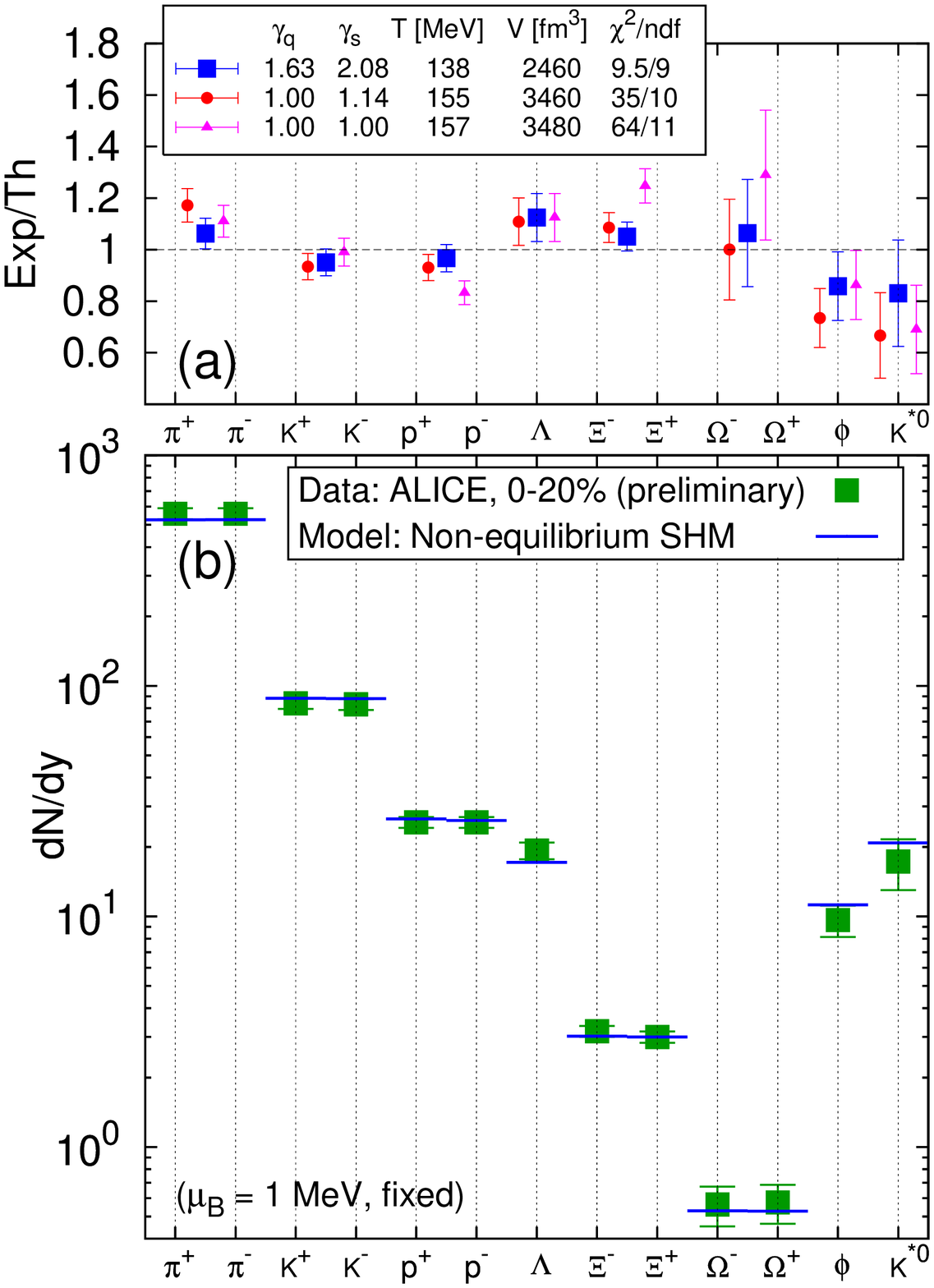}
\includegraphics[height=12.cm]{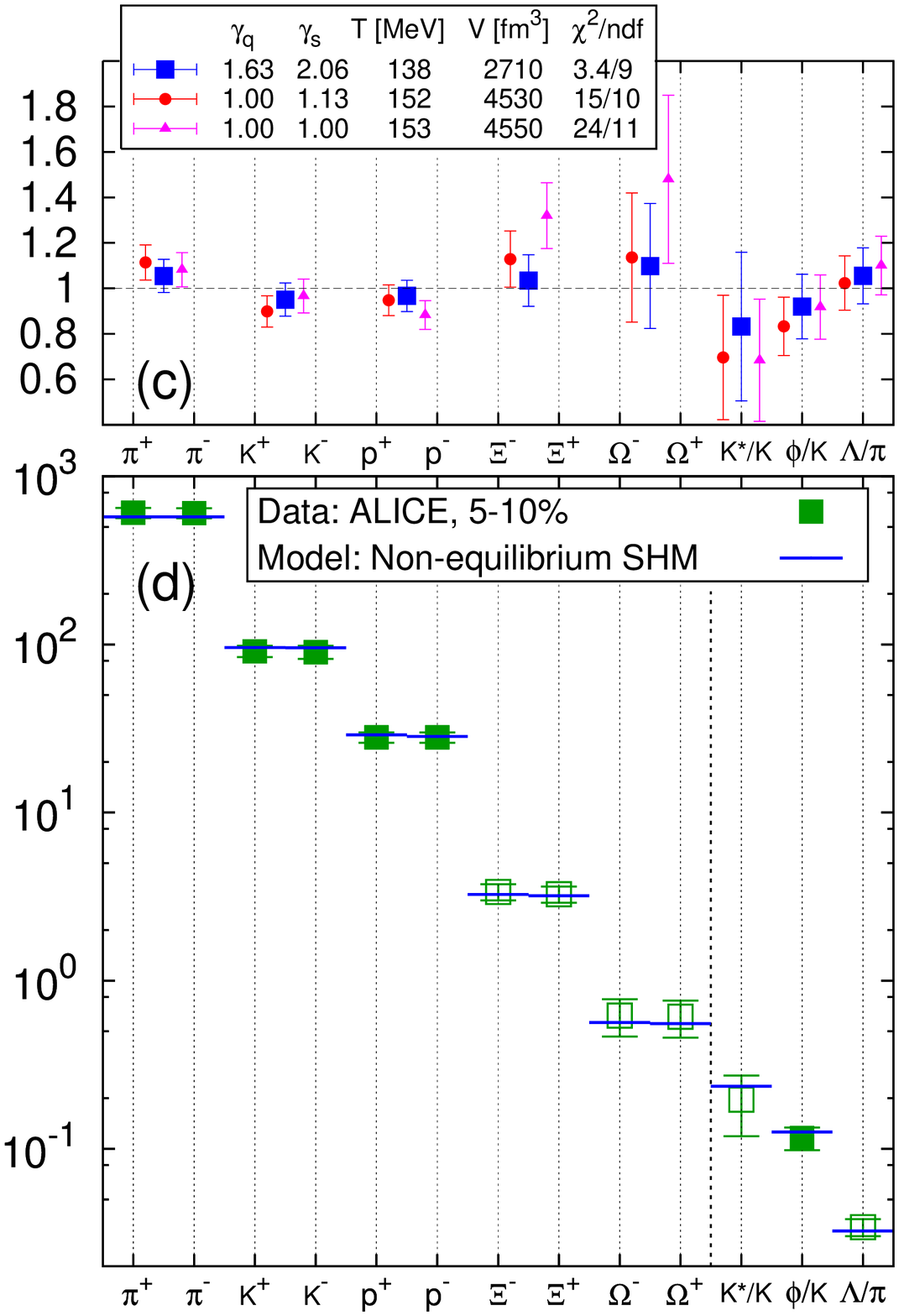}
\caption{\label{fig:firstfit}(color online)  {\small SHM} fit to  experimental data measured by the {\small ALICE} experiment  in Pb--Pb collisions at $\sqrt{s_{NN}}=2.76$ TeV for  0--20\% centrality (panels (a) and (b) on left hand side) and for 1/4 of this range, 5--10\% (panels (c) and (d) on right hand side). The input set of particle types is the same as can be seen in particle listing on the ordinate of panels (b) and (d), in panel (d) also particle yield ratios are used. In the lower panels (b) and (d) comparison of {\small SHM} chemical non-equilibrium fit (horizontal line) with data is shown. The experimental data is shown as filled  square, in the panel (d) the interpolated experimental data is shown with open symbols (see appendix~\ref{sec:data} for details). The upper panels (a) and (c) show the ratio of model values to experimental data for the three {\small SHM}  variants and present the key parameter values for: chemical non-equilibrium (solid  squares), chemical semi-equilibrium  (solid  circles) and chemical equilibrium (solid triangles).  For readability anti-particles are omitted in panels (a) and (c).}
\end{figure*}

The chemical non-equilibrium results for the 5--10\% centrality bin  (containing interpolated data, open symbols) is shown for comparison in the right panel of \rf{fig:firstfit}. The fit has the same set of particles as the 0--20\% centrality bin, however, we must fit here three ratios for which data are directly (or by interpolation) available, and we use a more recent set of proton, pion and kaon data. Definition of the model and some technical details about how we obtain results seen in \rf{fig:firstfit}  follow below; the fitted data are shown in the fourth column of table~\ref{tab:data}. The \rf{fig:firstfit}c  shows  the {\small SHM} parameters and $\chi^2$ for all three variants.    Comparing the {\rm SHM} parameters on left and right-hand of \rf{fig:firstfit} we see a large change in   $V$ expected for different centralities. We see that use of finer centrality binning and more mature data sample  reduces $\chi^2$ for all {\small SHM} variants.

As figure~\ref{fig:firstfit} shows and we discuss below in detail,  the chemical non-equilibrium {\small SHM} works perfectly at {\small LHC}, resulting in  a high  confidence level. This could be predicted considering prior CERN Super Proton Synchrotron ({\small SPS}) and {\small RHIC} data analyses~\cite{Rafelski:2004dp,Letessier:2005qe,Petran:2011aa}, which  strongly favor chemical non-equilibrium variant of {\small SHM}. Moreover, the chemical non-equilibrium {\small SHM} has a dynamical physical foundation in sudden breakup of a {\small QGP} fireball, we are not aware of a dynamical origin of the simple chemical equilibrium {\small SHM} since no dynamical computation of relativistic heavy ion scattering  achieves the chemical equilibrium condition without introduction of unknown particle, cross sections, etc.. Furthermore, as we will discuss in subsection \ref{sec:physicalproperties}, we obtain hadronization universality across a wide collision energy range: comparing {\small RHIC}62 with {\small LHC}2760 we show that the fireball source of particles is nearly identical, and consistent with  chemically equilibrated {\small QGP} fireball. Given this result,  chemical non-equilibrium  {\small SHM} variant is validated across a wide energy range, while  the chemical equilibrium {\small SHM}~\cite{Becattini:2012sq,BraunMunzinger:2001ip,Becattini:2010sk,Becattini:2005xt,BraunMunzinger:2003zd,Andronic:2009gj} is invalidated by the {\small LHC} data and this conclusion can be extended across different reaction energies  as there is no reason why a model should work only sporadically. 

We have now shown that the chemical non-equilibrium is the necessary ingredient in the {\small SHM}  approach to the process of hadronization of a {\small QGP} fireball.   The non-equilibrium {\small SHM} was proposed  when first strange hadron multiplicity results  were interpreted more than 20 years ago~\cite{Rafelski:1991rh}.  The yield of strange hadrons indicated that the number of  quark pairs present had to be modified by a factor $\gamma_s$,  the source of strangeness is not populating the final state hadrons  with the yields expected from the hadronic   chemical equilibrium,  a point of view widely accepted today. At {\small SPS} energies, for which this model was originally conceived, production of strangeness did not yet saturate the {\small QGP} phase space, that is strangeness was out of chemical equilibrium   both in {\small QGP} fireball source with  $\gamma_s^{\rm Q}<1$,   and thus also in the final hadronic state with also $\gamma_s^{\rm H}<1$. The distinction of {\small QGP} as initial and the final hadron phase space domain  for $\gamma_s$ was also modeled~\cite{Rafelski:1996hf}. It is important to always remember  that hadron phase space non-equilibrium can arise from a {\small QGP} fireball with strangeness in chemical equilibrium, since, in general, the {\small QGP} and hadron phase space strangeness density are greatly different. Moreover, it is quite possible that a not yet in chemical equilibrium {\small QGP}, which is the higher density phase, produces an equilibrated hadron yield. This can, however, happen only accidentally and variation of reaction energy or collision centrality shows this. 

Another non-equilibrium parameter $\gamma_c$, similar  to $\gamma_s$, was  introduced very soon after $\gamma_s$ to control the charm final state phase space~\cite{Rafelski:1996hf}, and it has been widely adopted in consideration of a strong charm yield overabundance above chemical hadron gas equilibrium. Note that both strangeness and charm flavors are therefore assumed  to have been produced in a separate and independent process before hadronization --- and note further that each of the production mechanisms, in this case, is different with charm originating in first parton collisions and strangeness being also abundantly produced in secondary thermalized gluon fusion reactions. At the end of {\small QGP} expansion, these available and independently established strangeness and charm particle supplies are distributed into available final state phase space cells, that is the meaning of {\small SHM} in a nutshell.

The full chemical non-equilibrium  is introduced by means of the  parameter $\gamma_q\ne 1$. This situation arises when the source of hadrons disintegrates faster than the time necessary to re-equilibrate the yield of light quarks present. The two pion correlation data provide  experimental evidence that favors a rapid breakup of {\small QGP} with a short time of hadron production~\cite{Csernai:2002sj}, and thus favors very fast, or sudden, hadronization~\cite{Csernai:1995zn,Rafelski:2000by}. In this situation, a similar chemical non-equilibrium approach must be applied to the light quark abundance, introducing the light quark phase space occupancy $\gamma_q$. This proposal made for the high energy {\small SPS} data~\cite{Letessier:1998sz,Letessier:2000ay}, helped improve the understanding of {\small RHIC}200 hadron rapidity yield results~\cite{Rafelski:2004dp} and allowed a consistent interpretation of these data across the full energy range at {\small SPS} and {\small RHIC}200~\cite{Letessier:2005qe}.  

For more than a decade we have made continued effort to show that a high quality (low $\chi^2$)  and simple (no need for hybrid models) description of hadron abundances emerges using chemical non-equilibrium SHM. However, the {\em recognition} of the necessity of light quark $(u,d)$ chemical non-equilibrium, i.e., $\gamma_q\ne 1$, remains  sparse, despite consistency of this approach with  the two pion correlation results which provides additional  evidence for fast hadronization~\cite{Csernai:2002sj}. The recent steady advances of lattice {\small QCD}~\cite{Endrodi:2011gv,Bazavov:2011nk,Philipsen:2012nu,Borsanyi:2012rr} favors {\small QGP} hadronization at a temperature  below the once preferred $T_c = 165\,\mathrm{MeV}$ temperature. As already noted the equilibrium  {\small SHM} variant imposing $\gamma_q=1$ light quark chemical equilibrium~\cite{Becattini:2012sq,BraunMunzinger:2001ip,Becattini:2010sk,Becattini:2005xt,BraunMunzinger:2003zd,Andronic:2009gj} produces (relatively dense) particle chemical freeze-out near to $T=155\,\mathrm{MeV}$. Such freeze-out assumes on one hand in the present context a relatively  high QGP hadronization temperature, and on the other hand requires as a complement an `afterburner' describing further reaction evolution of some particles. As we will argue in section \ref{sec:pOverPi}, such `hybrid' model does not result in a viable description of the precise {\small ALICE} experimental data.

This is the case since the {\small LHC}2760 experimental environment has opened  a new experimental opportunity to investigate in detail  the {\small SHM} hadron production model.  Precise particle tracking near to interaction vertex in the {\small ALICE} experiment removes the need for off-line corrections of weak interaction decays, and at the same time vertex tracking is  enhancing the efficiency of track identification,  increasing considerably the precision of particle yield measurement~\cite{Abelev:2012vx,Ivanov:2013haa}.  All data used in the present work was obtained in this way  by the {\small LHC-ALICE} experiment  for Pb--Pb collisions at $\sqrt{s_{NN}}=2.76$ TeV,   limited to the  central unit of rapidity interval $-0.5<y<0.5$. The experimental particle yield results are reported  in different collision centrality bins according to the geometric overlap of colliding nuclei, with the `smallest', e.g., 0--5\% centrality bin corresponding to the nearly fully overlapping geometry of the colliding nuclei. Collision geometry model~\cite{Abelev:2013qoq} relates the centrality trigger to the number of participating nucleons $N_{\rm part}$ which we use as our preferred centrality variable in what follows.

Section~\ref{sec:parameters} presents our general method and approach to the particle multiplicity data analysis. Following a brief summary  of the {\small SHM} methods in subsection~\ref{ssec:general}, we describe  in subsection~\ref{sec:CentralityStudy}   our centrality study of particle production based on the following  data: for the 0--20\% centrality bin, we obtain the preliminary data from~\cite{Abelev:2012vx,Ivanov:2013haa}. For the centrality study of particle production, we present in table~\ref{tab:data} the final yields of $\pi^\pm$, K$^\pm$ and p$^\pm$ as presented in~\cite{Abelev:2013vea}. The preliminary ratio  $\phi/$K, is from~\cite{Singha:2012qv}; these 7 data points are binned in the same centrality bins and are used as presented. However, several other particle types require rebinning with interpolation and, at times, extrapolation  which is further discussed  in appendix~\ref{sec:data}. The (preliminary) data input into this rebinning    for K$^{*0}/$K$^-$  and for ${2\Lambda}/{(\pi^-\!\!+\pi^+)}$  are also taken from Ref.~\cite{Singha:2012qv}. Using the preliminary enhancement factors of $\Xi^-$, $\Xibar^+$, $\Omega^-$, $\Omegabar^+$  shown in Ref.~\cite{Chinellato:2012jj,Knospe:2013ir}, combined with  yields of these particles for p--p-reactions at $\sqrt{s_{NN}}=7$ TeV as presented in Ref.~\cite{Abelev:2012jp}, we obtain  the required yield input, see appendix~\ref{sec:data}.   In subsection~\ref{sec:particleY}, we present particles both fitted and predicted by {\small SHM}, including  anti-matter clusters. 

In section~\ref{sec:Fireball}, we discuss the key physics outcome of the fits, i.e., the resulting {\small SHM} parameters as a function of centrality. We compare to the equilibrium approach  in subsection~\ref{sec:thermalproperties}. We discuss the differences seen between the {\small SHM} variants and compare our results to our analysis of Au--Au collisions at $\sqrt{s_{NN}}=62.4\GeV$ at {\small RHIC}62, as it is a system we analyzed in detail recently~\cite{Petran:2011aa}. We obtain the bulk physical properties: energy density, entropy density, and pressure, as a function of centrality in subsection~\ref{sec:physicalproperties}, where we also address strangeness and entropy yields. This  study is made   possible since all {\small SHM} parameters are determined with minimal error in consideration of the precise experimental particle multiplicity result. We  discuss how our results relate to the lattice-{\small QCD} study of {\small QGP} properties in subsection~\ref{sec:lattice}. We close our paper with a short summary and discussion of all results in section~\ref{sec:conclusion}.

\section{\label{sec:parameters}{\small SHM} and particle production}
\subsection{\label{ssec:general}Generalities}

We use here {\small SHM} implementation within the {\small SHARE} program~\cite{Torrieri:2004zz}. The {\small SHM} describes the yields of particles given the chemical freeze-out temperature $T$ and overall normalization $dV/dy$ (as the experimental data are available as $dN/dy$).  We account for the small asymmetry between particles and anti-particles by fugacity factors  $\lambda_q,\,\lambda_s$ and the light quark asymmetry  $\lambda_{I3}$, see Ref.\cite{Torrieri:2004zz}. We further note that it is not uncommon to present the particle--anti-particle asymmetry employing the baryo-chemical and strangeness chemical potentials defined by
\begin{equation}\label{chempot}
\mu_B =3T\ln \lambda_q\quad \text{and}\quad \mu_S=T\ln (\lambda_q/\lambda_s),
\end{equation}
the `inverse' definition of $\mu_S$ with reference to $\lambda_s$ has historical origin and is source of frequent error.

For each value of  $\lambda_q$, strangeness fugacity $\lambda_s$ is evaluated by imposing the strangeness conservation requirement $\langle s\rangle-\langle \bar s\rangle\simeq 0$. From now on, we omit the bra-kets indicating grand canonical average of the corresponding summed particle yield. The isospin fugacity factor $\lambda_{I3}$ is constrained by imposing the charge per baryon ratio present in the initial nuclear matter state at initial instant of the collision. We achieve this objective by fitting these conservation laws along with particle yield data, using  the following form:
\begin{align}
\frac{s-\overline{s}}{s+\overline{s}} &= 0.00\pm0.01,  \label{eq:convS}\\
\frac{Q-\overline{Q}}{B-\overline{B}} &= 0.38\pm0.02. \label{eq:convQ}\\
\end{align}
We believe that implementing conservation laws as data points with errors accounts for possibility that particles  escape asymmetrically from the acceptance domain. 

In the {\small LHC}2760 energy regime, there is near symmetry of particle and anti-particle sector thus the chemical potentials are hard to quantify. Therefore, the two constraints \req{eq:convS} and \req{eq:convQ}  alone were not sufficient to achieve smooth behavior of the chemical potentials as a function of centrality.  We therefore impose as a further constraint a constant baryon number stopping per participating nucleon in the mid-rapidity region in the following form:
\begin{equation}
\frac{b-\overline{b}}{N_{\rm part}} = 0.0054\pm 1\%. \label{eq:netbaryon}
\end{equation}
We selected condition \req{eq:netbaryon} since this was the variable which emerged in unconstrained fits as being most consistent. The value we selected is our estimate based on convergence without constraint to this value at several centralities. The alternative to this approach would have been to take a constant value of $\mu_B$  across centrality. While this produces a good enough fit as well, this approach was poorly motivated: the unconstrained fit results produced rather random looking distribution of  $\mu_B$   across centrality and thus did not present any evidence pointing towards  a specific choice for  $\mu_B$. While the actual method of fixing matter--anti-matter asymmetry is extraneous to the main thrust of this paper, the value of $\mu_B$ is of some relevance when considering predictions for anti-nuclei which we present further below.

Our considerations include the already described phase space occupancy parameters $\gamma_s$ and $ \gamma_q$, where the light quarks $q=u,d$ are not distinguished. We do not study $\gamma_c$ here, in other words, we do not include in present discussion the charm degree of freedom. We note that there is no current experimental $p_\bot$-integrated charmed hadron yield information available from Pb--Pb collisions at {\small LHC}.  The integration of the phase space distribution is not yet possible due to uncertain low transverse momentum yields. 

Thus, in  {\small LHC}2760 energy domain, we have at most $4=7-3$ independent statistical model parameters--constraints: seven parameters $dV/dy, T, \lambda_q, \lambda_s, \lambda_{I3}, \gamma_q $ and $\gamma_s$ constrained by the three conditions, \req{eq:convS}, \req{eq:convQ} and \req{eq:netbaryon}, to describe within {\small SHM} approach many very precise data points spanning in yield across centrality more than 5 orders of magnitude. We will show  for comparison results obtained setting arbitrarily $\gamma_q=1$ (chemical semi-equilibrium fit, comprising $6-3$ parameters--constraints) and than $\gamma_q=\gamma_s=1$ (chemical  equilibrium fit, $5-3$ parameters--constraints).

Absolute yields of hadrons are proportional to one power of $\gamma_q$ for each constituent light quark (or anti-quark) and one power of  $\gamma_s$ for each strange quark (or anti-quark). For example, $\gamma_q$ enters non-strange baryon to meson ratios in the following manner:
\begin{equation}
\label{RatioBM}
\frac{{\rm baryon}(qqq)}{{\rm meson}(q\overline{q})}\propto\frac{\gamma_q^3}{\gamma_q^2}F(T, m_{\rm baryon}, m_{\rm meson}),
\end{equation}
where $q$ stands for either $u$ or $d$ quark and $F $ is the integral over all particle momenta of the phase space distribution at  freeze-out temperature: we always use exact form of  relativistic phase space integrals. For  strange hadrons, we must replace $\gamma_q$ by $\gamma_s$ for each constituent $s$ (and/or $\overline{s}$) quark. Experimentally measured light baryon to meson ratios (such as p$/\pi$) strongly depend on the value of $\gamma_q$ in a fit. Similarly, $\Lambda(qqs)/\pi(q\overline{q})\propto\gamma_s$ is very sensitive to the value of $\gamma_s$.

The value of $\gamma_q$ is bound by appearance of a pion condensate which corresponds to a singularity in the pion Bose--Einstein  distribution function reached at the condition
\begin{equation}
\gamma_q^\mathrm{crit} = \mathrm{exp}\left(\frac{m_{\pi^0}}{2T}\right).
\label{eq:gammacrit}
\end{equation}
This numerically works out for $T=138$--$160$ MeV to be in range $\gamma_q^\mathrm{crit}=1.63$--$1.525$. On the other hand, there is a  much more lax limit on the range of $\gamma_s$, strangeness can increase very far before a particle condensation phenomenon limit is reached for the $\eta$ meson.

\subsection{\label{sec:CentralityStudy}Centrality study}
\begin{table*}[t]\scriptsize
\caption{\label{tab:data} Table of data points we use as input for {\small SHM} fits; header of the table defines the centrality bins in three different ways.  Hadron yields ($dN/dy$) and ratios at mid-rapidity $|y|<0.5$ for different centralities. Centrality as a function of $N_\mathrm{\rm part}$ taken over from~\cite{Abelev:2013qoq}. Errors are combined systematic and statistical errors added in quadratures where systematic errors are in general dominant and statistical errors are negligible. Values in brackets are interpolated data. See appendix~\ref{sec:data} for details about data sources and how data are rebinned.}
\resizebox{\textwidth}{!} {
\begin{ruledtabular}
\begin{tabular}{l|c|ccccccccc}
Centrality & 0--20\%&0--5\% & 5--10\% & 10--20\% & 20--30\% & 30--40\% & 40--50\% & 50--60\% & 60--70\% & 70--80\%  \\
$\langle N_\mathrm{\rm part}\rangle$& 308 & 382.8 & 329.7 & 260.5 & 186.4 & 128.9 & 85.0 & 52.8 & 30.0 & 15.8 \\
$dN_{\rm ch}/d\eta$ & & ${1601\pm60}$ & ${1294\pm49}$ & ${966\pm37}$ & ${649\pm23}$ & ${426\pm15}$ & ${261\pm9}$ & ${149\pm6}$ & ${76\pm4}$ & ${35\pm2}$ \\
\hline
$\pi^+$ & ${562\pm36}$ & ${733\pm54}$ & ${606\pm42}$ & ${455\pm31}$ & ${307\pm20}$ & ${201\pm13}$ & ${124\pm8}$ & ${71\pm5}$ & ${37\pm2}$ & ${17.1\pm1.1}$ \\
$\pi^-$ & ${560\pm34}$ & ${732\pm52}$ & ${604\pm42}$ & ${453\pm31}$ & ${306\pm20}$ & ${200\pm13}$ & ${123\pm8}$ & ${71\pm4}$ & ${37\pm2}$ & ${17.0\pm1.1}$  \\
K$^+$ & ${84\pm5.4}$ & ${109\pm9}$ & ${91\pm7}$&${68\pm5}$&${46\pm4}$& ${30\pm 2}$&${18.3\pm 1.4}$&${10.2\pm 0.8}$&${5.1\pm 0.4}$&${2.3\pm 0.2}$\\
K$^-$ & ${84\pm5.7}$ & ${109\pm9}$ & ${90\pm8}$&${68\pm6}$&${46\pm4}$& ${30\pm 2}$&${18.1\pm 1.5}$&${10.2\pm 0.8}$&${5.1\pm 0.4}$&${2.3\pm 0.2}$\\
K$^{*0}$ & ${17.3\pm4.2}$ & \\
K$^{0*}/$K$\, 10^3$ &  &$(188\pm98)$ & $(196\pm77)$ & $(209\pm54)$ & $(227\pm59)$ & $(247\pm64)$ & $(269\pm70)$ & $(295\pm77)$ & $(326\pm85)$ & $(361\pm94)$ \\
$\mathrm{p}$ &${25.9\pm1.6}$& ${34\pm 3}$ & ${28\pm 2}$ & ${21.0\pm 1.7}$ & ${14.4\pm 1.2}$ & ${9.6\pm 0.8}$ & ${6.1\pm 0.5}$ & ${3.6\pm 0.3}$ & ${1.9\pm 0.2}$ & ${0.90\pm 0.08}$ \\
$\mathrm{\overline{p}}$ &${26.0\pm1.8}$& ${33\pm 3}$ & ${28\pm 2}$ & ${21.1\pm 1.8}$ & ${14.5\pm 1.2}$ & ${9.7\pm 0.8}$ & ${6.2\pm 0.5}$ & ${3.7\pm 0.3}$ & ${2.0\pm 0.2}$ & ${0.93\pm 0.09}$  \\
$\phi$    & ${9.6\pm1.4}$ & \\
$\phi/$K$\, 10^3$ & & ${109\pm20}$ & ${116\pm18}$ & ${117\pm17}$ & ${128\pm19}$ & ${120\pm20}$ & ${123\pm18}$ & ${123\pm19}$ & ${119\pm18}$ & ${119\pm21}$ \\ 
$\Lambda$ & ${19.3\pm2.0}$ & \\
$\Lambda/\pi\, 10^3$ & & $(33.3\!\pm\!3.9)$ & $(34.2\!\pm\!4.0)$ & ${35.3\!\pm\!4.1}$ & $(36.4\!\pm\!4.3)$ & $(37.0\!\pm\!4.3)$ & $(37.1\!\pm\!4.4)$ & $(36.8\!\pm\!4.3)$ & $(36.0\!\pm\!4.2)$ & $(34.7\!\pm\!4.1)$   \\
$\Xi^-\, 10^2$    & ${323\pm35}$& $(397\pm44)$ & $(337\pm37)$ & $(258\pm28)$ & $(176\pm19)$ & $(116\pm13)$ & $(71.6\pm7.9)$ & $(40.7\pm4.5)$ & $(19.6\pm2.2)$ & $(7.5\pm0.8)$  \\ 
$\Xibar^+\, 10^2$    & ${304\pm33}$& $(382\pm42)$ & $(327\pm36)$ & $(253\pm28)$ & $(176\pm19)$ & $(118\pm13)$ & $(73.7\pm8.1)$ & $(42.3\pm4.7)$ & $(20.3\pm2.2)$ & $(7.1\pm0.8)$ \\ 
$\Omega^-\, 10^2$ &${57\pm10}$& $(78\pm19)$ & $(62\pm16)$ & $(45\pm11)$ & $(29\pm7)$ & $(18\pm4)$ & $(10\pm3)$ & $(5.4\pm1.4)$ & $(2.5\pm0.6)$ & $(1.0\pm0.3)$   \\ 
$\Omegabar^+\, 10^2$ &${58\pm11}$& $(76\pm18)$ & $(61\pm15)$ & $(45\pm11)$ & $(29\pm7)$ & $(18\pm4)$ & $(10\pm3)$ & $(5.5\pm1.4)$ & $(2.5\pm0.6)$ & $(1.0\pm0.3)$ \\ 
\end{tabular}
\end{ruledtabular}
} 
\end{table*}

The input hadron yield data used in the fit to the  0--20\% centrality bin is shown in the 2nd column of table \ref{tab:data}. The   fit to this data set for the case of chemical equilibrium, where one forces  $\gamma_s=\gamma_q=1$,   was done in Ref.~\cite{Ivanov:2013haa} choosing a fixed value $\mu_B=1$\,MeV. In a first step, we compare to these results and   follow  this approach. However, we consider it necessary to apply strangeness and charge per baryon conservation by fitting \req{eq:convS} and \req{eq:convQ} as two additional data points determining the corresponding values of chemical parameters $\mu_S, \mu_{I3}$, a procedure  omitted in the report Ref.~\cite{Ivanov:2013haa} where   $\mu_S=\mu_{I3}=0$ was set. Naturally, the effect of this improvement is minimal, but it assures physical consistency. We show the values of  $\chi^2_\mathrm{total}$   in \rf{fig:chi2}, see the large open symbols. The wider range of $N_\mathrm{\rm part}$ corresponding to the centrality bin 0--20\%  is shown in \rf{fig:chi2} as a horizontal uncertainty bars.

\begin{figure}[!tb]
\includegraphics[width=0.96\columnwidth]{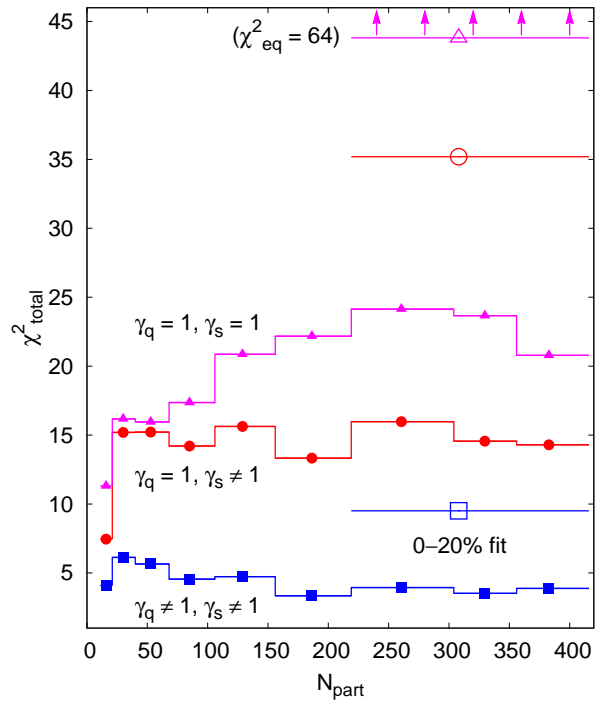}
\caption{\label{fig:chi2}(color online) Total $\chi^2$  as a function of centrality, as indicated in figure, for the total equilibrium ($\gamma_q=\gamma_s=1,\,\mathrm{ndf}=11$), for the semi-equilibrium ($\gamma_q=1,\,\gamma_s\neq 1,\,\mathrm{ndf}=10$), and for chemical non-equilibrium ($\gamma_q\neq 1,\,\gamma_s\neq 1,\,\mathrm{ndf}=9$) {\small SHM}. Open symbols represent the total $\chi^2$  for the   0--20\% centrality bin depicted in figures \ref{fig:firstfit}a,b The number of degrees of freedom for the three cases, respectively, is $\mathrm{ndf}=11,10,9$. (The value of equilibrium SHM in the 0--20\% bin has been shifted down by 20 in order to fit in the figure.)}
\end{figure}

In our detailed centrality dependent analysis, we use data in nine finer centrality bins, which we show in the third to eleventh and last column of table~\ref{tab:data}. The bins are classified according to the average number of participants $N_{\rm part}$ as a measure of centrality. This is  a model value originating in the experimentally measured pseudo-rapidity density of charged particles $dN_{\rm ch}/d\eta$~\cite{Abelev:2013qoq}, which we state in the third row of table~\ref{tab:data}. We consider the consistency   in \rf{fig:ncharged}: the experimentally measured $dN_{\rm ch}/d\eta$ in the relevant participant bins~\cite{Abelev:2013vea} is shown by square symbols, as well as our {\small SHM} results for rapidity density of charged particles $dN_{\rm ch}/dy$ emerging directly from {\small QGP} (i.e., primary charged hadrons) and the final yield of charged hadrons, as triangles,  fed by the decay of  hadronic resonances.  In all cases, we show, in \rf{fig:ncharged}, the yield per pair of interacting nucleons using the model value $N_{\rm part}$. While the primary charged hadron rapidity yield (full circles) is well below the pseudo rapidity density $dN_\mathrm{ch}/d\eta$ of charged hadrons (full squares), the final rapidity yield  $dN_\mathrm{ch}/dy$ after strong decays (full triangles) is well above it. This result, $dN_\mathrm{ch}/dy>dN_\mathrm{ch}/d\eta$ is consistent with dynamical models  describing the momentum spectra, which are accounting for production of charged particles that are not identified by experiments~\cite{Bozek:2012qs}.

\begin{figure}[!thb]
\includegraphics[width=\columnwidth]{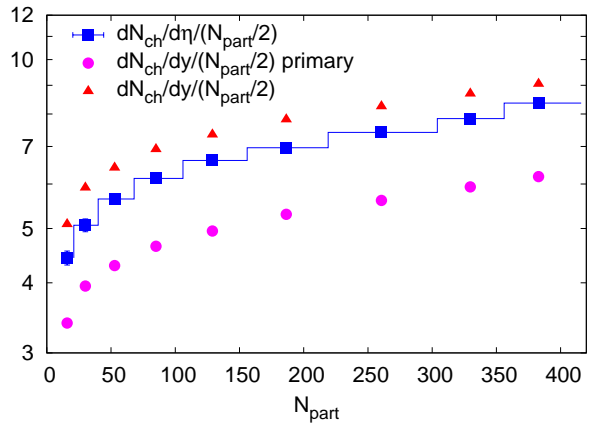}
\caption{\label{fig:ncharged}(color online) Experimental charged particle yield pseudo-rapidity density $dN/d\eta$ (blue squares), and geometric model relating to charged particle rapidity density $dN/dy$ of only primary particles (violet circles) and including feed from strongly decaying resonances (red triangles) per participant pairs $N_{\rm part}/2$.}
\end{figure}

In \rf{fig:ncharged}, we see that about 50\% of charged hadronic particles are  produced by strong decays of heavier resonances. We show, in \rf{fig:primary_ratios}, the ratio of primarily produced yield to the total  yield for different particle species in the expected range of hadronization temperatures. The dominant fraction, almost 80\%, of $\pi$ and $\mathrm{p}$ yield originates from decaying resonances. This result demonstrates the difficulty that one encounters in the interpretation of transverse momentum spectra which must account for the decays and is thus, in a profound way, impacted by collective flow properties of many much heavier hadrons~\cite{Bozek:2012qs,Rybczynski:2012ed,Broniowski:2001we}. Conversely, this means that  one can perform a convincing analysis of transverse momentum distribution  only for hadrons, which do not have a significant feed from resonance decays, such as $\Omega$ or $\phi$. This finding is   the   reason  why we study the $p_\bot$ integrated yields of hadrons in exploration of the physics of the fireball particle source.  Moreover, we believe that  `blast-wave' model fits to $p_\bot$ hadron spectra are only meaningful for the $\Omega$ or $\phi$ hadrons.

\begin{figure}[!t]
\includegraphics[width=\columnwidth]{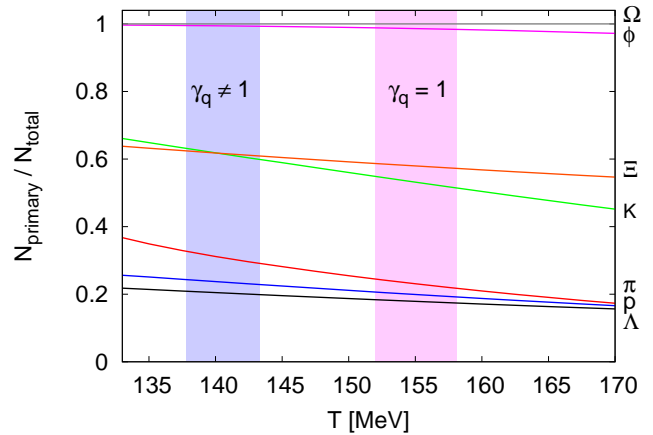}
\caption{\label{fig:primary_ratios}(color online) Fraction of primary hadrons produced, normalized by  their final yield, which consists of primary produced hadrons and the feed from strong decaying resonances, for particles indicated on the right margin, as a function of hadronization temperature in single-freeze-out model.}
\end{figure}

The centrality binning, which differs for different particles considered, requires us to use several interpolated and even some slightly extrapolated experimental results, which procedure we discuss in depth below, and in appendix~\ref{sec:data}. In our fits, we choose to use the centrality bins with the largest number of directly determined experimental data, minimizing the potential error originating in our multi-point interpolation. A few particles appear more than once in our data set (as a yield and/or in a ratio). However, to prevent duplicity, we always fit every particle measured just once.

In order to show that the finer centrality binning matters, we have already shown the 5--10\% centrality bin  (which  contains only close extrapolation, see open symbols in the figures \ref{fig:firstfit}b,d). The fit has the same set of particles as the 0--20\% centrality bin seen in figures \ref{fig:firstfit}a,c, however, some of these particles enter the finer binned fit in ratios. If the outcome of the fit as a function of centrality is even a small variation in fitted parameters (other than normalization, i.e., volume), we expect and we find that the 5--10\% centrality bin, which describes a much smaller participant ${N_\mathrm{\rm part}}$ range, leads to  smaller $\chi^2$ compared to the wide 0--20\% case. However, the stability of the fit parameters implies that much of the improvement is  attributable to the revision in the input data set. The 0--20\% fit is based on preliminary data~\cite{Ivanov:2013haa}, whereas 5--10\% includes more recent final data~\cite{Abelev:2013vea} (see appendix A for details). For chemical non-equilibrium SHM an improvement of $\chi^2$ by a factor of 4 is found for both preliminary 0--20\% and more recent final data set in 5--10\% bin as compared to chemical equilibrium SHM, thus favoring our simple non-equilibrium hadronization model.

We perform a fit to the entire data set with all three {\small SHM} approaches and compare the resulting  $\chi^2$ as a function of ${N_\mathrm{\rm part}}$ in \rf{fig:chi2}. The solid squares represent the chemical non-equilibrium {\small SHM} ($\gamma_q\neq 1,\gamma_s\neq 1$), the solid circles represent the semi-equilibrium {\small SHM}  ($\gamma_q=1,\gamma_s\neq1$) and solid triangles represent the full equilibrium {\small SHM}   ($\gamma_q=\gamma_s=1$). The range of centrality is indicated by the horizontal bars. 
Considering most central bins, we note in \rf{fig:chi2} that allowing $ \gamma_q\ne 1$ can reduce the total $\chi^2$ of the fit by more than a factor of $3$ compared to semi-equilibrium, and more than a factor of $5$ comparing full non-equilibrium with full equilibrium. 

As a last step, we verify if there is a special value of the parameter $\gamma_q$ of particular importance. To this end,  we have evaluated the $\chi^2/\mathrm{ndf}$ of the fit as a function of a given fixed $\gamma_q$    within a range  $\gamma_q\in (0.95,\gamma_q^\mathrm{crit})$. This $\chi^2$ profile curves, seen in \rf{fig:chi-profiles},  all pass  $\gamma_q=1$ smoothly,  therefore $\gamma_q=1$ has no special importance for the {\small SHM}. However, fits to data in all centralities decrease in $\chi^2$ as $\gamma_q$ increases, they all point to best fit value of $\gamma_q$ near the critical value of Bose--Einstein  condensation   given by \req{eq:gammacrit}.

The most peripheral bin (70--80\%, $N_{\rm part}=15.8$) analyzed here  requires further discussion  as it shows in \rf{fig:chi-profiles} a  different behavior and in particular a considerably lower $\chi^2$ when $\gamma_q\to 1$. For this peripheral centrality bin the procedure we use to interpolate data of $\Xi$, $\Omega$, $\Lambda/\pi$ and K$^*$/K  assigns a narrow peripheral centrality range to these experimental data points obtained for a much greater centrality domain  spanning a participant range which is considerably wider. This can be a problem  since within the wider centrality range the experimental results change  rapidly with participant number. Therefore, our extrapolation towards the edge of the experimental data centrality range may introduce a fit aberration, here it happens that the created data are less incompatible with equilibrium SHM variants when $\gamma_q\to 1$. We do not believe that there is any issue with result of the fit for  $\gamma_q\to 1.6$ we discuss in this work. A different approach, in which we recombine the bins rather than to inter-extrapolate was presented in Ref.\cite{Petran:2013qla}.

\begin{figure}[!t]
\includegraphics[width=\columnwidth]{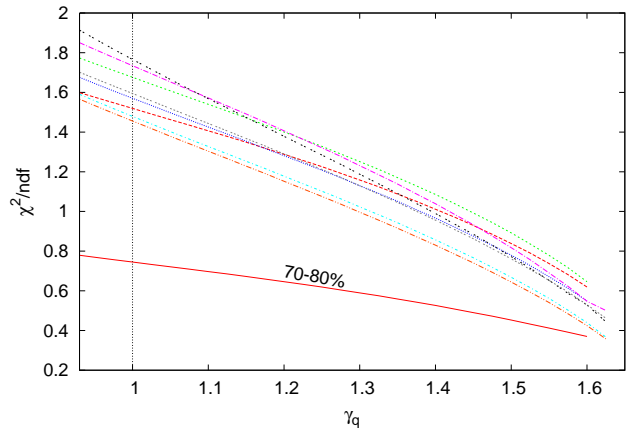}
\caption{\label{fig:chi-profiles}(color online) $\chi^2/\mathrm{ndf}$ profile as a function of $\gamma_q$ for all studied centralities.}
\end{figure}

\subsection{\label{sec:particleY}Particle yields}
We compare input to the  resulting particle yields graphically in \rf{fig:fittedparticles}. We fit 13 particles, counting anti-particles, which in the figure cannot be visually distinguished as an independent input  or output data, and ratios $\Lambda/\pi$, K$^*$/K and $\phi/$K. For these ratios,  the relevant yield outputs $\Lambda$, K$^*$ and $\phi$ are shown. Direct comparison of the input $\Lambda/\pi$, K$^*/$K and $\phi/$K ratios to the output is presented in \rf{fig:KstarLambdaRatios}, note that $\phi/$K ratio is available as experimental data point in all centrality bins. The  fitted  output yields are stated also in the top portion of table~\ref{tab:yields1}, and ratios are given just below allowing for comparison with the input values.

\begin{figure}[!t]
\centerline{\includegraphics[width=0.95\columnwidth]{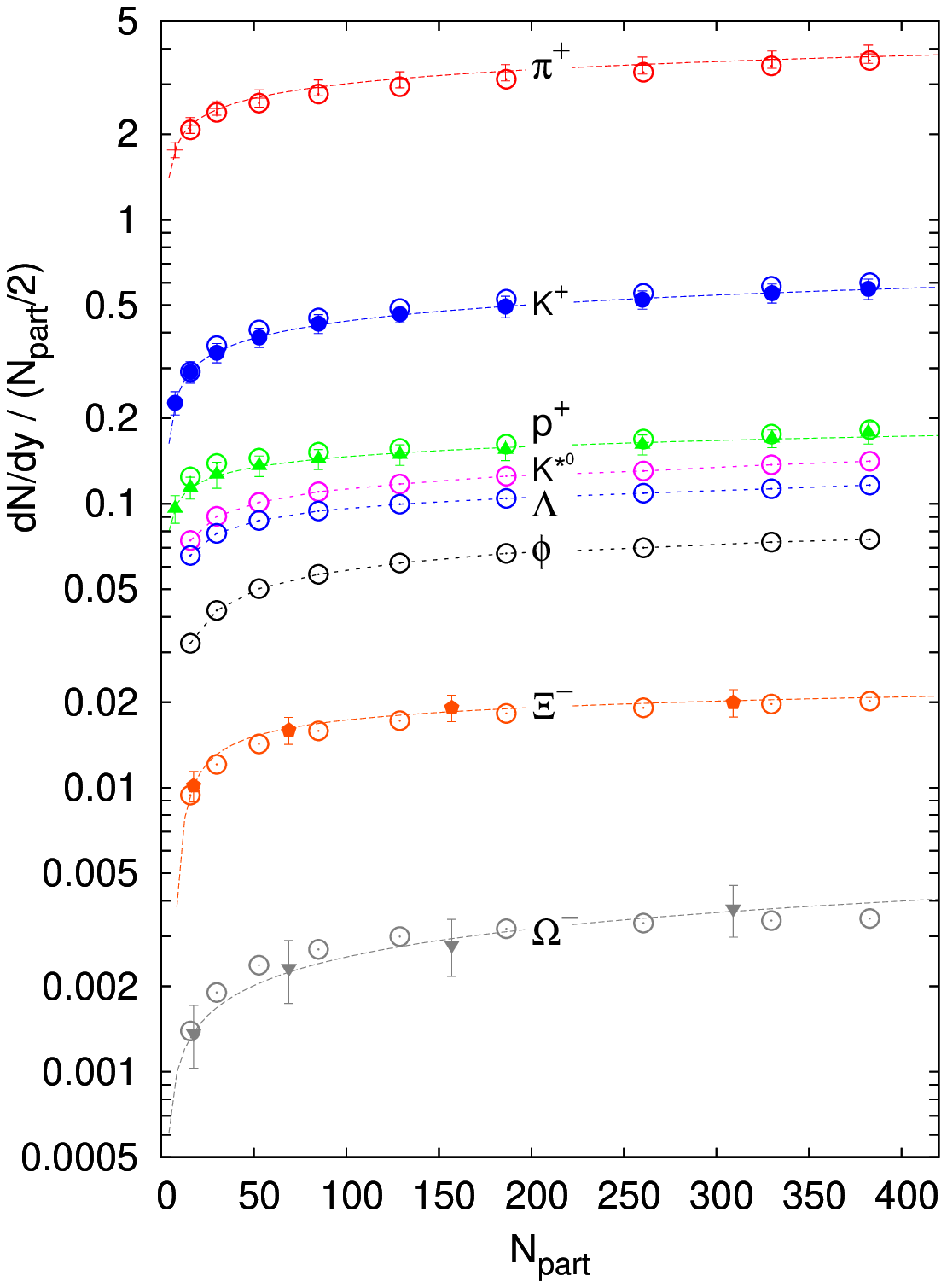}}
\caption{\label{fig:fittedparticles}(color online) Full symbols are the experimental data points. Open symbols represent the outcome for our  chemical non-equilibrium {\small SHM} fit to {\small LHC}2760 as a function of centrality, i.e., $N_{\rm part}$. Dashed lines are outcome of our data interpolation of experimental yields available (see appendix~\ref{sec:data}). Dotted lines connecting $\Lambda$, K$^*$ and $\phi$ {\small SHM} output values are presented  to guide the eye; since ratios of these particles were  used in our fit, the data are shown in \rf{fig:KstarLambdaRatios}.}
\end{figure}

\begin{table*}[!ht]
\caption{\label{tab:yields1} Table of hadron yield output; header of the table defines the centrality bins in three different ways. Top section of the table shows fitted yields $dN/dy$ of hadrons entering the fit at {\small LHC}2760 obtained in  the chemical non-equilibrium {\small SHM}. Next below  are  three ratios that are actually included in the fit (rather than the yields of $\Lambda,$ K$^{0*}$,$\phi$), followed by  the ratios of hadron yields that can  be formed from the stated results, stated here for convenience of the reader. In the two lower sections of the table, there are  predicted yields of yet unmeasured hadrons and at the very bottom, we show predicted yields of light anti-nuclei scaled up by factor 1000 (and  by $10^6$ for anti-Helium). Note that yield of matter particles is nearly the same.}
\begin{ruledtabular}
\resizebox{\textwidth}{!}{ 
\begin{tabular}{l|c|ccccccccc}
Centr.bin &0--20\%& 0--5\% & 5--10\% & 10--20\% & 20--30\% & 30--40\% & 40--50\% & 50--60\% & 60--70\% & 70--80\%       \\ 
$\langle N_\mathrm{\rm part}\rangle$& 308 & 382.8   & 329.7    & 260.5     & 186.4     & 128.9     & 85.0      & 52.8      & 30.0        & 15.8 \\ 
$dN_{\rm ch}/dy$ & $1312$ & $1732$ & $1433$ & $1075$ & $729$ & $474$ & $294$ & $169$ & $88.7$ & $40.2$\\
\hline
$\pi^+$ & $525$ & $696$ & $574$ & $431$ & $292$ & $190$ & $118$ & $68.0$ & $35.8$ & $16.4$\\
$\pi^-$ & $525$ & $696$ & $575$ & $430$ & $292$ & $189$ & $118$ & $68.1$ & $35.9$ & $16.4$\\

K$^+$ & $88.4$ & $115$ & $96.0$ & $71.8$ & $49.0$ & $31.4$ & $19.2$ & $10.8$ & $5.41$ & $2.30$\\
K$^-$ & $88.1$ & $114$ & $95.5$ & $72.1$ & $48.7$ & $31.5$ & $19.1$ & $10.8$ & $5.37$ & $2.30$\\

$\mathrm{p}$ & $26.5$ & $34.9$ & $29.0$ & $22.0$ & $15.1$ & $10.1$ & $6.45$ & $3.82$ & $2.08$ & $0.982$\\
$\mathrm{\overline{p}}$ & $26.1$ & $34.2$ & $28.4$ & $21.6$ & $14.8$ & $9.89$ & $6.30$ & $3.72$ & $2.02$ & $0.953$\\

K$^{0*}$ & $20.8$ & $27.0$ & $22.6$ & $17.0$ & $11.7$ & $7.56$ & $4.68$ & $2.67$ & $1.36$ & $0.587$\\

$\phi$ & $11.2$ & $14.4$ & $12.1$ & $9.12$ & $6.23$ & $3.99$ & $2.40$ & $1.33$ & $0.632$ & $0.255$\\

$\Lambda$ & $17.2$ & $22.3$ & $18.6$ & $14.2$ & $9.73$ & $6.43$ & $4.00$ & $2.31$ & $1.18$ & $0.520$\\

$\Xi^-$ & $3.03$ & $3.86$ & $3.25$ & $2.49$ & $1.70$ & $1.11$ & $0.674$ & $0.377$ & $0.181$ & $0.0745$\\
$\Xibar^+$ & $3.00$ & $3.80$ & $3.22$ & $2.43$ & $1.68$ & $1.08$ & $0.664$ & $0.371$ & $0.179$ & $0.0726$\\

$\Omega^-\,10^3$ & $529$ & $663$ & $561$ & $435$ & $297$ & $193$ & $115$ & $62.6$ & $28.5$ & $11.0$\\
$\Omegabar^+\,10^3$ & $527$ & $654$ & $560$ & $421$ & $295$ & $186$ & $114$ & $62.1$ & $28.4$ & $10.7$\\
\hline
$\Lambda/\pi\,10^3$ & $32.7$ & $32.0$ & $32.4$ & $33.0$ & $33.4$ & $33.9$ & $34.0$ & $33.9$ & $32.9$ & $31.7$\\
K$^{0*}\!/$K$\,10^3$ & $236$ & $235$ & $237$ & $236$ & $239$ & $240$ & $245$ & $248$ & $252$ & $255$\\
$\phi/$K$\,10^3$ & $127$ & $125$ & $126$ & $127$ & $128$ & $127$ & $125$ & $123$ & $117$ & $111$\\
\hline
$\phi/\pi^-\,10^3$ & $21.4$ & $20.6$ & $21.0$ & $21.2$ & $21.4$ & $21.1$ & $20.3$ & $19.5$ & $17.6$ & $15.6$\\
K$^-/\pi^-\,10^3$ & $168$ & $165$ & $166$ & $167$ & $167$ & $167$ & $162$ & $158$ & $150$ & $141$\\
p$/\pi^+\,10^3$ & $50.4$ & $50.2$ & $50.5$ & $51.0$ & $51.8$ & $53.2$ & $54.8$ & $56.2$ & $58.0$ & $60.0$\\
$\Xi/\pi\,10^3$ & $5.76$ & $5.55$ & $5.65$ & $5.79$ & $5.84$ & $5.86$ & $5.71$ & $5.53$ & $5.06$ & $4.55$\\
$\Omega/\pi\,10^3$ & $1.007$ & $0.952$ & $0.976$ & $1.010$ & $1.019$ & $1.019$ & $0.973$ & $0.920$ & $0.795$ & $0.671$\\
K$^{0*}\!/\pi\,10^3$ & $39.6$ & $38.8$ & $39.3$ & $39.5$ & $40.0$ & $39.9$ & $39.7$ & $39.2$ & $37.8$ & $35.9$\\
$\mathrm{p/K}$ & $0.300$ & $0.303$ & $0.302$ & $0.306$ & $0.309$ & $0.322$ & $0.336$ & $0.354$ & $0.385$ & $0.426$\\
\hline \hline
$\eta$ & $61.0$ & $79.7$ & $66.3$ & $49.8$ & $33.9$ & $21.9$ & $13.4$ & $7.64$ & $3.90$ & $1.72$\\
$\rho(770)^0$ & $38.9$ & $51.3$ & $42.5$ & $32.1$ & $21.9$ & $14.5$ & $9.14$ & $5.35$ & $2.87$ & $1.34$\\
$\omega(782)^0$ & $35.1$ & $46.4$ & $38.4$ & $29.0$ & $19.8$ & $13.1$ & $8.28$ & $4.85$ & $2.61$ & $1.22$\\
$\Delta(1232)^{++}$ & $4.98$ & $6.57$ & $5.46$ & $4.15$ & $2.86$ & $1.92$ & $1.23$ & $0.734$ & $0.402$ & $0.191$\\
$\Sigma^*(1385)^{-}$ & $2.08$ & $2.70$ & $2.26$ & $1.72$ & $1.18$ & $0.785$ & $0.492$ & $0.284$ & $0.146$ & $0.065$\\
$\Lambda^*(1520)$ & $1.09$ & $1.41$ & $1.18$ & $0.907$ & $0.625$ & $0.418$ & $0.264$ & $0.153$ & $0.0795$ & $0.0355$\\
$\Xi^*(1530)^{-}$ & $1.02$ & $1.30$ & $1.09$ & $0.84$ & $0.58$ & $0.378$ & $0.230$ & $0.129$ & $0.0626$ & $0.0258$\\
\hline \hline
$^2\mathrm{\overline{H}}\,10^3$ & $74.7$ & $98.1$ & $81.9$ & $62.6$ & $43.6$ & $29.4$ & $19.5$ & $11.8$ & $6.53$ & $3.16$\\
$^3_\Lambda\mathrm{\overline{H}}\,10^3$ & $0.478$ & $0.601$ & $0.506$ & $0.397$ & $0.279$ & $0.191$ & $0.128$ & $0.0773$ & $0.0415$ & $0.0193$\\
$^3\mathrm{\overline{H}}\,10^3$ & $1.64$ & $2.13$ & $1.79$ & $1.39$ & $0.983$ & $0.677$ & $0.468$ & $0.290$ & $0.166$ & $0.083$\\
$^3\mathrm{\overline{He}}\,10^3$ & $1.64$ & $2.14$ & $1.79$ & $1.39$ & $0.986$ & $0.679$ & $0.469$ & $0.291$ & $0.166$ & $0.083$\\
$^4\mathrm{\overline{He}}\,10^6$ & $5.87$ & $7.57$ & $6.41$ & $5.04$ & $3.64$ & $2.56$ & $1.85$ & $1.18$ & $0.697$ & $0.362$\\
\end{tabular}
} 
\end{ruledtabular}
\end{table*}

Our fit results appear as open circles in \rf{fig:fittedparticles}, at times completely overlaying the input data, full symbols. For the $\Lambda$, the dotted line guides the eye, since the actual fit is to the ratio $\Lambda/\pi$ shown in \rf{fig:KstarLambdaRatios}, no absolute $\Lambda$ data are available, in absence of absolute yields, only open circle, i.e., the fitted value, is shown in  \rf{fig:fittedparticles}.  Similar situation arises with $\phi$ and K$^*$, where data are not available, but we fit $\phi/$K and K$^{0*}/$K. One can see that {\small SHM} generated results follow closely both the experimental data available, and the interpolation  dashed lines  for each particle, and that each interpolation curve passes  through the experimental data points shown in full symbols,  or at worse, the error bars if these are larger than the symbol. 

\begin{figure}[!t]
\centerline{\includegraphics[width=0.97\columnwidth]{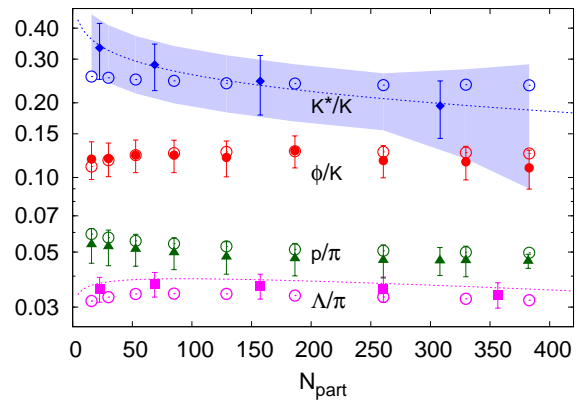}}
\caption{\label{fig:KstarLambdaRatios}(color online) Full symbols: experimental data  with errors for K$^*/$K, $\phi/$K, and $\Lambda/\pi$ as a function of centrality, {i.e.}, number of participants $N_{\rm part}$. Lines are the interpolation, respectively, extrapolation except for the case $\phi/$K, where dotted line guides the eye. Open circles represent the resulting fit value for each ratio. The (blue) shaded band shows the error input used obtaining the interpolated values of K$^*/$K ratio.}
\end{figure}

Even so, we note in \rf{fig:fittedparticles}, that our  interpolation for $\Omega$ shows a slightly different systematic shape (dashed line) compared  the fit results (open symbols) or the behavior of the other particles. In other words, we see that other particles `predict' the yield of  $\Omega$  that follows the centrality dependence of other particles, while the four data points lead  to a centrality distribution that is slightly different.  More precise $\Omega$  data will without any doubt offer a resolution to this slight tension in our interpolation/extrapolation.   The hadron yields we find are also  stated in  table~\ref{tab:yields1}.  Aside from the yields, we show there frequently quoted ratios of particle yields, e.g., we find  p$/\pi^+\simeq 0.05$. We will return to discuss this ratio in subsection \ref{sec:pOverPi}.

Figure~\ref{fig:KstarLambdaRatios} has the largest  differences between theory and experiment.  In case of the   $\Lambda/\pi$ ratio, we see a systematic within error bar  under-prediction at all centralities.  For K$^*/$K, we see within the error bar a different slope of the fit as a function of centrality. The question can be asked if these  differences of fit and results indicate some not yet understood physics contents. However, we are within error bars and such data--fit difference must be expected and is allowed given a large data sample and potential for experimental refinement of these two preliminary data sets involving K$^*$ and $\Lambda$. We recall that at {\small RHIC}200, K$^*/$K ratio was 10--15\% smaller and agrees with current {\small ALICE} results within the error margin~\cite{Aggarwal:2010mt}.  We also note  that we did not yet study how the charmed hadron decay particles influence the fit.

\begin{figure}[!t]
\centerline{\includegraphics[width=0.95\columnwidth]{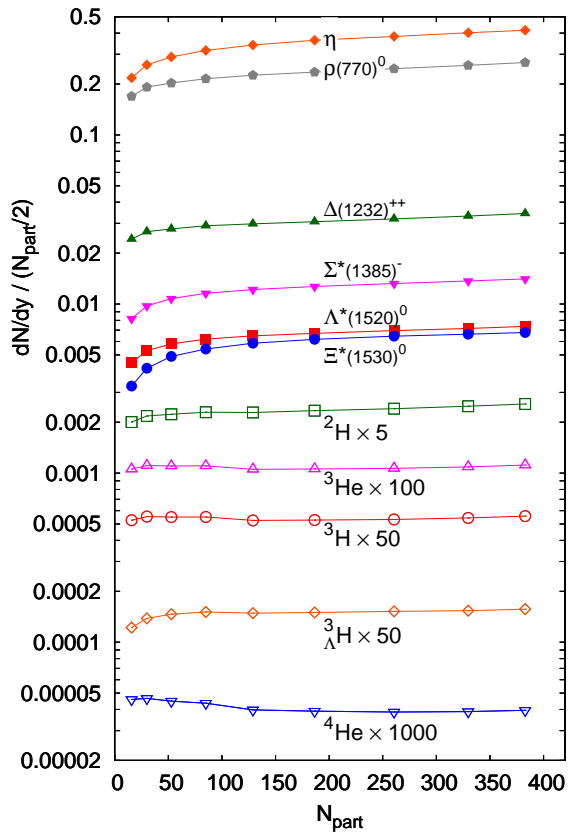}}
\caption{\label{fig:predictedparticles}(color online) Particles predicted by the chemical non-equilibrium {\small SHM} fits. Predictions for hadron yields (full symbols) are above anti-nuclei yield predictions (open symbols), which have been multiplied by suitable factors. Lines  guide the eye, points are actual predictions for each of the nine centralities we analyzed.}
\end{figure}

Predictions for the six hadron yields $\eta$, $\rho^0$, $\Delta(1232)^{++}$, $\Lambda^*(1520)$, $\Sigma^*(1385)^{-}$, $\Xi^*(1530)^{0}$,  are shown in \rf{fig:predictedparticles} as a function of centrality, these results are  stated in the lower portion of table~\ref{tab:yields1}. We further show five different species of (strange) anti-matter, from anti-deuteron to anti-alpha, including anti-hypertriton, appropriately scaled to fit into display of \rf{fig:predictedparticles}. Our predictions of these composite objects should serve as a lower limit of their production rates: fluctuation in the  {\small QGP} homogeneity at hadronization, and recombinant  formation after hadronization may add   contributions to the small {\small SHM} yield, see here the corresponding {\small RHIC} result~\cite{Abelev:2010rv}.

\section{\label{sec:Fireball}Particle source and its properties}

\subsection{\label{sec:thermalproperties}Statistical parameters}
In  \rf{fig:thermalparameters}, we depict the   {\small LHC}2760 statistical parameters as a function of collision centrality and compare these {\small LHC}2760 results with those we have obtained at {\small RHIC}62~\cite{Petran:2011aa}, shown with open symbols.  In all three panels of \rf{fig:thermalparameters}, we show parameter errors evaluated by {\small SHARE}v2~\cite{Torrieri:2004zz} employing the {\small MINOS} minimization routine. One can see that the parameter values for chemical non-equilibrium are defined better than for the case with $\gamma_q=1$.

\begin{figure}[!hbt]
\includegraphics[width=\columnwidth]{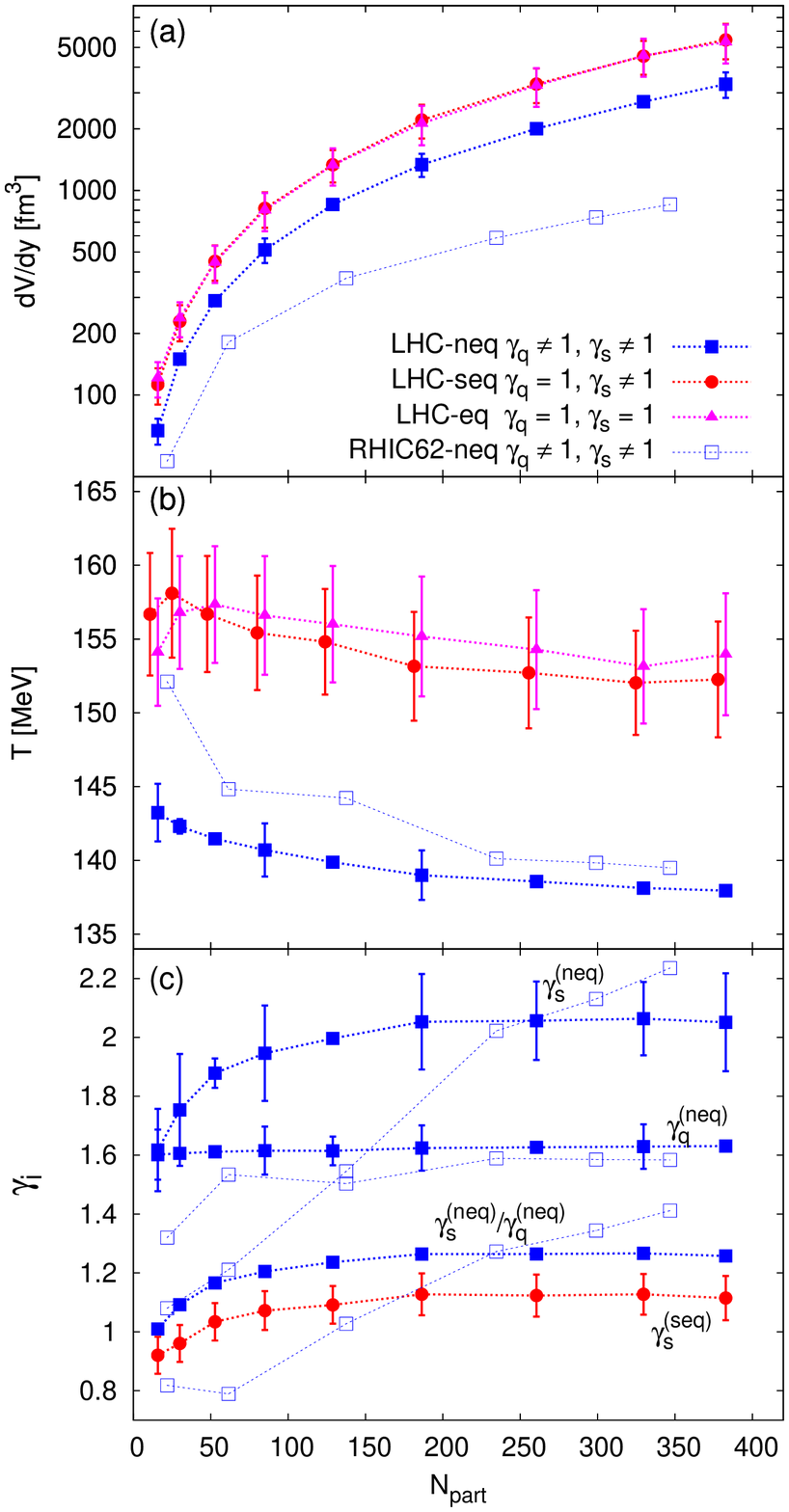}
\caption{\label{fig:thermalparameters} (color online) {\small SHM} parameters as a function of centrality, i.e., number of participants $N_{\rm part}$, presented for the three  different levels of chemical equilibrium, and compared to the chemical non-equilibrium {\small SHM} {\small RHIC}62 results (open symbols, dashed lines). All lines guide the eye. From top to bottom: (a)   $dV/dy$ --- note that the volume for both equilibrium and semi-equilibrium {\small SHM} variants is so close that symbols overlap; (b)   $T$, the chemical freeze-out temperature (semi-equilibrium symbols are offset in order to separate them from equilibrium); (c)  phase space occupancies $\gamma_s^{(\mathrm{neq})}$, $\gamma_q^{(\mathrm{neq})}$ and for comparison with  equilibrium $\gamma_s^{(\mathrm{seq})}$; we also present $\gamma_s^{(\mathrm{neq})}/\gamma_q^{(\mathrm{neq})}$.}
\end{figure}

We present  {\small LHC}2760 hadronization parameters for  the non-equilibrium {\small SHM} case also in the top section of table~\ref{tab:params1}. In the top frame \rf{fig:thermalparameters}a,  we see the particle source volume $dV/dy$, in the middle frame \rf{fig:thermalparameters}b,  the chemical freeze-out temperature $T$, and in the bottom frame \rf{fig:thermalparameters}c, the phase space occupancies --- the different variants are distinguished by superscripts `neq' (non-equilibrium, that is $\gamma_q\neq 1, \gamma_s\neq 1$) , `seq' (semi-equilibrium,  $\gamma_q=1, \gamma_s\neq 1$).   and `eq' (equilibrium,  $\gamma_q=1, \gamma_s= 1$).  To compare with the  semi-equilibrium {\small SHM} variant, we show the ratio $\gamma_s^{(\mathrm{neq})}/\gamma_q^{(\mathrm{neq})}$, a ratio which helps to quantify the strangeness to light quark enhancement. This is to be directly compared with the  semi-equilibrium  strangeness phase space occupancy $\gamma_s^{(\mathrm{seq})}$, given fixed $\gamma_q^{(\mathrm{seq})}=1$. 

\begin{table*}[!htb]\footnotesize
\caption{\label{tab:params1}Top section shows chemical non-equilibrium {\small SHM} fit parameters $dV/dy$, $T$, $\gamma_q$, $\gamma_s$ and $\chi^2_{\rm total}$ with ndf (number data less number of parameters) obtained in each centrality bin. For error discussion see text in section~\ref{sec:physicalproperties}. Bottom section presents fireball bulk properties in each bin: energy density $\varepsilon$, pressure $P$, entropy density $\sigma$, strangeness per entropy content $s/S$, entropy at {\small LHC}2760 compared to {\small RHIC}62, $S_{\mathrm{LHC}}/S_{\mathrm{RHIC}}$ and net baryon number per entropy ratio $b/S$}.
\begin{ruledtabular}
\resizebox{\textwidth}{!}{ 
\begin{tabular}{l|c|ccccccccc}
Centrality  & 0--20\%& 0--5\% & 5--10\% & 10--20\% & 20--30\% & 30--40\% & 40--50\% & 50--60\% & 60--70\% & 70--80\%   \\
$\langle N_\mathrm{part}\rangle$& 308 & 382.8   & 329.7    & 260.5     & 186.4     & 128.9     & 85.0      & 52.8      & 30.0       & 15.8\\
\hline\\[-0.2cm]
$dV/dy\,\mathrm{[fm^3]}$ & $2463\!\pm\!6$  & $3304\!\pm\!469$  & $2715\!\pm\!81$  & $2003\!\pm\!47$  & $1337\!\pm\!173$  & $853.9\!\pm\!5.9$ & $512.2\!\pm\!70.1$ & $289.4\!\pm\!5.5$ & $149.8\!\pm\!5.0$ & $66.9\!\pm\!9.7$\\[0.07cm]
$T\,\mathrm{[MeV]}$      & $138.3\!\pm\!0.0$ & $138.0\!\pm\!0.0$ & $138.1\!\pm\!0.0$ & $138.6\!\pm\!0.0$ & $139.0\!\pm\!1.7$ & $139.9\!\pm\!0.0$ & $140.7\!\pm\!1.8$ & $141.5\!\pm\!0.0$ & $142.3\!\pm\!0.5$ &$143.2\!\pm\!2.0$\\[0.07cm]
$\gamma_q$               & $1.63\!\pm\!0.00$ & $1.63\!\pm\!0.00$ & $1.63\!\pm\!0.08$ & $1.63\!\pm\!0.00$ & $1.62\!\pm\!0.08$ & $1.62\!\pm\!0.05$ & $1.62\!\pm\!0.08$ & $1.61\!\pm\!0.00$ & $1.61\!\pm\!0.00$ &$1.60\!\pm\!0.09$\\[0.07cm]
$\gamma_s$               & $2.08\!\pm\!0.00$ & $2.05\!\pm\!0.17$ & $2.06\!\pm\!0.13$ & $2.06\!\pm\!0.13$ & $2.05\!\pm\!0.16$ & $2.00\!\pm\!0.01$ & $1.95\!\pm\!0.16$ & $1.88\!\pm\!0.05$ & $1.75\!\pm\!0.19$ &$1.62\!\pm\!0.14$\\[0.07cm] 
$\chi^2_{\rm total}/{\rm ndf}$       & $9.51/9$ & $3.87/9$ & $3.52/9$ & $3.94/9$ & $3.35/9$ & $4.73/9$ & $4.55/9$ & $5.65/9$ & $6.13/9$& $4.09/9$ \\[0.07cm] 
\hline\\[-0.2cm]
$\varepsilon\,\mathrm{[GeV/fm^3]}$ & $0.462$ & $0.453$ & $0.457$ & $0.467$ & $0.476$ & $0.487$ & $0.505$ & $0.516$ & $0.521$& $0.527$ \\[0.04cm]
$P\, \mathrm{[MeV/fm^3]}$          & $78.5 $ & $77.1 $ & $77.7 $ & $79.1 $ & $80.5 $ & $82.3 $ & $85.1 $ & $86.8 $ & $87.9 $& $89.2 $ \\[0.04cm]
$\sigma\, \mathrm{[fm^{-3}]}$      & $3.20 $ & $3.14 $ & $3.17 $ & $3.23$ & $3.28 $ & $3.36 $ & $3.46 $ & $3.53 $ & $3.56 $& $3.60 $ \\[0.04cm]
$s/S$                    & $0.0299$ & $0.0295$ & $0.0297$ & $0.0297$ & $0.0298$ & $0.0294$& $0.0290$ & $0.0284$ & $0.0272$ & $0.0257$ \\[0.04cm]
$S_{\mathrm{LHC}}/S_{\mathrm{RHIC}}$     & $3.07$ & $3.23$ & $3.10$ & $2.93$ & $2.75$ & $2.56$ & $2.33$ & $2.06$ & $1.74$& $1.27$ \\
$b/S\times 10^4$				& $1.37$			& $2.00$ & $2.08$ & $2.19$ & $2.30$ & $2.45$ & $2.61$ & $2.81$ & $3.06$ & $3.57$\\
\end{tabular}
} 
\end{ruledtabular}
\end{table*}

For the  {\small LHC}2760 data, the {\small SHM} forcing chemical equilibrium of light quarks (i.e., $\gamma_q=1$ with either $\gamma_s =1$ or $\gamma_s\ne 1$) have a very similar volume $dV/dy$, and similar chemical freeze-out $T$  as shown in \rf{fig:thermalparameters}a,b, respectively, with nearly overlapping lines for $dV/dy$. In the non-equilibrium approach, $dV/dy$ is reduced by about 20--25\%, and the freeze-out temperature $T$ by 10\% compared to the equilibrium {\small SHM} variant. Compared to the {\small RHIC}62 results~\cite{Petran:2011aa}  (open symbols) the {\small LHC}2760 volume $dV/dy$ is up to a factor 4 larger while the {\small LHC} hadronization temperature $T$ is 2--5 MeV lower. Thus, given equal number of participants $N_\mathrm{part}$ at {\small RHIC}62 and {\small LHC}2760, the much larger particle multiplicity $dN/dy$ requires in consideration of the universal hadronization condition~\cite{Petran:2013qla} considerably increased transverse dimension of the fireball at the time of hadronization, which we find within our {\small SHM} interpretation of hadron production data. We understand this growth of particle multiplicity (and therefore volume) as being due to a greater transverse  fireball expansion, driven by the greater initial energy density formed in {\small LHC}2760 heavy ion collision. This corresponds to a greater initial pressure necessary for the matter expansion to the same bulk hadronization conditions as already found  at {\small RHIC}. The small but systematic decrease of the freeze-out temperature at {\small LHC}2760 compared to {\small RHIC}62 may be an indication of a greater supercooling caused by the  more dynamical {\small LHC} expansion.

The freeze-out temperature $T$ at {\small LHC}2760 decreases   when considering more central collisions, see \rf{fig:thermalparameters}b. In the hadronization scenario used in this work, this can be interpreted as being due to a deeper supercooling of the  most central and most energetic collision systems. We can extrapolate the freeze-out temperature to $N_\mathrm{\rm part}=0$ in the figure to set an upper limit on hadronization temperature at {\small LHC}2760, $T_{\mathrm{had}} \to 145\pm4\MeV$, applicable to a small (transverse size) fireball. This, then, is the expected hadronization temperature without supercooling.  
Excluding, in \rf{fig:thermalparameters}b,  the most peripheral $T$-fit point for {\small RHIC}62, which does not have a good confidence level, we see that $T$ at  {\small RHIC}62 converges towards the same maximum value as we found at {\small LHC}2760, thus confirming the determination of $T_{\mathrm{had}}$ as the common hadronization temperature without supercooling.

We show the phase space occupancies  $\gamma_q,\gamma_s$ in  \rf{fig:thermalparameters}c. We note that the  {\small LHC}2760 fit produces nearly a constant $\gamma_q$ as a function of centrality. However, $\gamma_s$ (and respectively $\gamma_s/\gamma_q $) decrease for more peripheral collisions towards unity suggesting that these flavors approach the same level of chemical equilibrium for systems of small transverse size. A similar situation for peripheral collisions was observed for {\small RHIC}62. However, at {\small RHIC}62, we see a strong centrality dependence of $\gamma_s$ and hence $\gamma_s/\gamma_q$. This rapid rise of the {\small RHIC}62  $\gamma_s$ as a function of centrality can be attributed to the buildup of strangeness in {\small QGP} formed at {\small RHIC}62, which is imaged in the later produced strange hadron yield. Note that, omitting  the most peripheral {\small RHIC}62 point, the  peripheral $\gamma_q$ is nearly the same as at {\small LHC}2760. The  small difference can be attributed to the smaller allowed value of  $\gamma_q$  for the slightly higher value of $T$ seen at  {\small RHIC}62.  

We have executed  all our fits allowing for the presence of the chemical potentials (\req{chempot}) characterizing the slight matter--anti-matter asymmetry present  at {\small LHC}2760. The  quality of the fit is not sufficiently improved including effectively one extra parameter ($\mu_{B}$, since $\mu_S$ is fixed by strangeness conservation)  to assure that the unconstrained results for $\mu_{B}$ are convincing. As mentioned in section~\ref{ssec:general}, we smooth the centrality dependence of $\mu_B$ by introducing baryon stopping fraction at mid-rapidity, that is imposing \req{eq:netbaryon} as an additional data point, a value that we saw a few times in the data without introducing this constraint.  This constraint leads to the  chemical potentials $\mu_B$ and $\mu_S$ presented in \rf{fig:muBmuS}, with the baryochemical potential $1 \le \mu_{B}\le 2.3 $ MeV and $\mu_S=0.0\pm0.5\,\mathrm{MeV}$ for all centralities, values an order of magnitude smaller than at {\small RHIC}62 and {\small RHIC}200. As we can see, even with the constraint, there are two centralities which do not agree with the trend set by the other seven data points. 

Data shown in \rf{fig:muBmuS} are not defined well enough to argue that we see a decrease of baryochemical potential with increasing centrality, since this outcome could be result of the bias we introduced. However, we think that for the most central collisions at {\small LHC}2760 there is some indication that  $\mu_{B}\simeq 1.5 $\,MeV. Dashed line in \rf{fig:muBmuS} indicates the resultant baryon per entropy, $b/S$, scaled with 5000, these values are also seen in table \ref{tab:params1}. This is a first estimate of this important result needed for comparison with the conditions prevailing in the big bang early Universe where $b/S\simeq 3.3\times 10^{-11}$~\cite{Fromerth:2012fe}.

\begin{figure}[t!]
\includegraphics[width=\columnwidth]{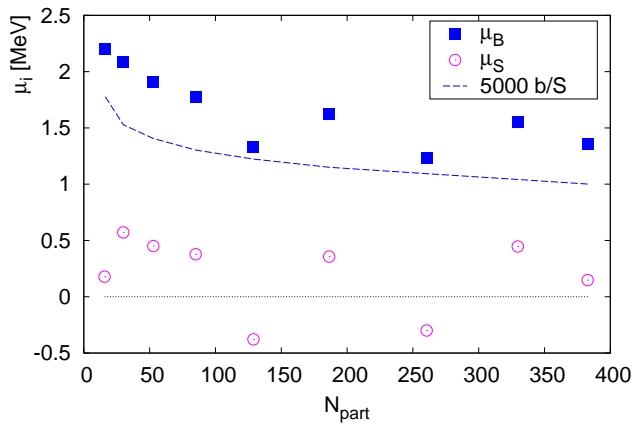}
\caption{\label{fig:muBmuS}(color online) Scatter plot of fitted chemical potentials. The dashed line shows netbaryon number over entropy $b/S$ scaled with 5000.}
\end{figure}

\subsection{\label{sec:pOverPi}p$/\pi$ ratio and chemical (non-)equilibrium}
The key difference between the three {\small SHM} approaches are the values of $\gamma_{q,s}$, as seen in \rf{fig:thermalparameters}c. In section~\ref{ssec:general}, we argued that the baryon to meson ratio, e.g., p$/\pi$, is directly proportional to $\gamma_q$ and this can be used to distinguish between the three {\small SHM} approaches. This ratio is a big problem for the equilibrium {\small SHM}~\cite{Abelev:2012wca}. We wish now to quantify this result within our approach and to show that, within the chemical non-equilibrium {\small SHM}, the problem is solved. 

For this purpose, we redo all fits but making this ratio more explicit in the data analysis. Specifically, first we evaluate p$/\pi$ ratio based on the yields of p and $\pi$ seen in table~\ref{tab:data} 
\begin{equation}
	\frac{\mathrm{p}}{\pi} \equiv \frac{\mathrm{p}\,\,+\,\,\overline{\mathrm{p}}}{\pi^-+\pi^+}.
\end{equation}
We estimate the error of the p$/\pi$ ratio by adopting the relative error of p$/\pi$ from~\cite{Abelev:2012wca}, that is 6.5\%. We include this new data point,  p$/\pi$ ratio,  in the fit. Note that this increases the relative importance of p and $\pi$ compared to the other particles included in the fit. Open symbols in the \rf{fig:pOverPi}a depict the data and full symbols show the resulting output values obtained when we refit with enlarged data set that includes the p$/\pi$ ratio. There is a minimal change in statistical parameters and physical properties of the fireball which we do not restate. In \rf{fig:pOverPi}b, we show  $\chi^2_{\rm total}$.

Even with the increased importance of p$/\pi$, the chemical non-equilibrium {\small SHM} works very well. However, {\small SHM} with fixed $\gamma_q=1$ have increased  difficulties describing this ratio, that is there is systematic 1.5--2 s.d. difference of the fit result and data and the value of  $\chi^2_{\rm total}$ is   large.  When compared to the $\chi^2_{\rm total}$ obtained without the added p$/\pi$ in \rf{fig:chi2}, the non-equilibrium variant shows nearly the same values of $\chi^2_{\rm total}$ for all centralities, the p$/\pi$ ratio is a natural outcome of the non-equilibrium approach. On the other hand, {\small SHM} approaches with $\gamma_q=1$  show additional  systematic increase in $\chi^2$ by a factor of $\sim 1.3$--1.5 for all centralities. This means that p$/\pi$ data are in conflict with the hypothesis $\gamma_q=1$.  This demonstrates that the hypothesis of chemical equilibrium of light quarks is incompatible with  the baryon to meson ratio at {\small LHC}2760 and $\gamma_q\simeq 1.6$ is needed in order to describe the {\small LHC} data. This finding is in agreement with the {\small RHIC}200 data~\cite{Huang:2005nda}, where the importance of the p$/\pi$ ratio was noted.

\begin{figure}[!bt]
\includegraphics[width=\columnwidth]{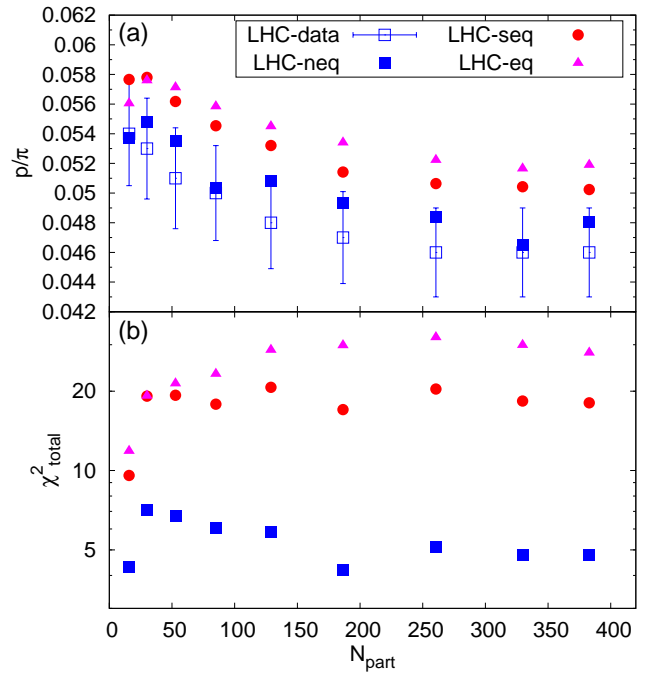}
\caption{\label{fig:pOverPi}(color online) Panel (a): data and {\small SHM} description of p$/\pi$ ratio fitted  together with all other data within the three {\small SHM} approaches, as a function of centrality. Panel (b): resulting $\chi^2_{\rm total}$ for the three variants. See section~\ref{sec:pOverPi} for details.}
\end{figure}
 
To compare p$/\pi$ ratio with our predictions, recall that the picture of universal hadronization condition with universal hadronization pressure $P=82\pm 5$\,MeV has been advanced by our group~\cite{Letessier:2005qe,Rafelski:2009jr,Petran:2013qla}. For this favored hadronization condition, the p$/\pi$ ratio is predicted in table II of Ref.~\cite{Rafelski:2010cw} to be p$/\pi = 0.047\pm 0.002$, which agrees practically exactly with the experimental result shown in \rf{fig:pOverPi}a. The {\small ALICE} collaboration~\cite{Abelev:2013vea} considers and discusses the mechanism of chemical equilibrium hadron production followed by post-hadronization interactions~\cite{Karpenko:2012yf,Rapp:2000gy,Pan:2012ne,Becattini:2012xb,Steinheimer:2012rd}, specifically proton--antiproton annihilation, in order to justify the small p$/\pi$ ratio, as compared to the result of equilibrium {\small SHM} alone. However, the annihilation mechanism was proposed based on preliminary data available in a single centrality bin 0--20\%, whereas our work includes more recent and centrality dependent experimental results~\cite{Abelev:2013vea}, allowing a far more conclusive study of the annihilation model.

Aside from $\mathrm{p\overline{p}}$ annihilation, there are $\mathrm{p\overline{p}}$ formation events. The significantly larger abundance (and therefore also density) of  heavy mesons compared to nucleons, see table~\ref{tab:yields1}, implies that mesons can be  an effective source of  nucleon pairs in reactions such as $\mathrm{p+\overline{p}} \longleftrightarrow \rho +\omega$ and many other relevant reactions, see table II in Ref.\cite{Amsler:1997up}. {\small ALICE} collaboration  notes, that   p$/\pi$ ratio modification after  annihilation  should disappear in most peripheral collisions due to smaller volume. We will now quantify  this effect  showing how this fade-out of the annihilation effect would work as a function of centrality. We show that given the constant p$/\pi$ ratio in a wide range of centralities,  \rf{fig:pOverPi}a, the effect of post-hadronization change of p$/\pi$ ratio must be negligible. 

In order to establish the  centrality dependence of post-hadronization nucleon yield changing reactions,  we  evaluate  the total number of  $\mathrm{p\overline{p}}$ annihilation events. This  number is obtained by integrating annihilation  rate over history of the post-hadronization matter expansion
\begin{equation}\label{annih}
N_{\rm annih}=\int  dt N_{\mathrm{\overline{p}}}(t)\rho_\mathrm{p}(t)\sigma_{\rm annih}v,
\end{equation}
where $v$ is the relative velocity of $\overline{p}$ and $p$. The three-dimensional dilution of the density can be modeled as
\begin{align}
\label{density_model}
\rho_\mathrm{p}(t)&=\frac{\rho_\mathrm{p}^h}{(1+\langle v_{\rm flow}  \rangle  t/\langle L \rangle)^3}, \\[0.2cm]
\langle L \rangle&\simeq [(dV/dy)/(4\pi/3)]^{1/3}, \nonumber
\end{align}
where $\langle L \rangle$ is the magnitude  of the fireball size, and $\langle v_{\rm flow} \rangle\simeq 0.6$--$1\,c$ is the velocity of the fireball expansion, in both cases averaged over the fireball complex three dimensional geometry. 

\begin{figure}[!b]
\includegraphics[width=\columnwidth]{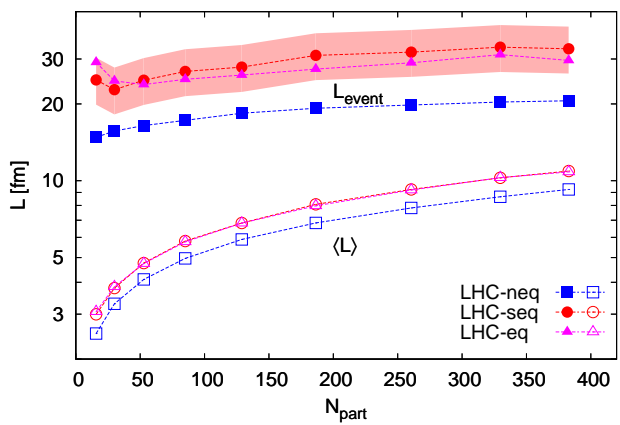}
\caption{\label{fig:proton-density}(color online) Full symbols: antiproton annihilation event  path $L_{\rm event}$ for the three {\small SHM} models as a function of centrality; open symbols: fireball size scale $\langle L \rangle$. Shaded error band represents error margin originating from the uncertainty of $T$ (semi)equilibrium SHM. For readability, we omit much smaller error band for  non-equilibrium SHM.}
\end{figure}

The initial density at time of hadronization is obtained from our hadronization study:
\begin{equation}\label{rho_had}
\rho_\mathrm{p}(t_h)\equiv \rho_\mathrm{p}^h \simeq \rho_{\mathrm{\overline{p}}}^h \equiv\frac{{dN_\mathrm{p}}/dy}{dV/dy}.
\end{equation}
In a wide range of low relative energies, which are relevant here, the event cross section is~\cite{Bruckner:1990aa}
\begin{equation}
\sigma_{\rm event}\equiv\sigma_{\rm annih} v/c\simeq 46\,\mathrm{mb}.
\end{equation}
Neglecting the depletion of nucleons (i.e., $N_{\mathrm{\overline{p}}}(t)\simeq N_{\mathrm{\overline{p}}}^h$),  we find, combining \req{annih} with \req{density_model}, the ratio of annihilated (anti)protons to their total yield ${N_{\rm annih}}/{N_{\mathrm{\overline{p}}}^h}$ and proton mean path before it annihilates $L_{\rm event}$:
\begin{equation}\label{annih-fraction}
\frac{N_{\rm annih}}{N_{\mathrm{\overline{p}}}^h}=  \frac{\langle L \rangle}{2 \langle v_{\rm flow}/c \rangle L_{\rm event}},\qquad  L_{\rm event}=\frac {1} {\sigma_{\rm event} \rho_\mathrm{p}^h}.
\end{equation}

The upper three lines, in \rf{fig:proton-density}, show  $ L_{\rm event}$ for the three models of hadronization (equilibrium, semi-equilibrium and non-equilibrium) as a function of centrality. The colored band in \rf{fig:proton-density} represents the error originating from the freeze-out temperature $T$ uncertainty (see \rf{fig:thermalparameters}). Note that the non-equilibrium model has much smaller parameter errors, so $L_{\rm event}$ is defined more precisely. Since the event reaction cross section for annihilation is well measured and nearly constant, it does not introduce any additional uncertainty to $L_{\rm event}$. The bottom three lines, in \rf{fig:proton-density} (semi-equilibrium and equilibrium lines overlap, since $dV/dy$ is very similar in these two cases),  show how the size $\langle L \rangle$ of the system changes as a function of centrality. Especially for peripheral collisions, we see that $\langle L \rangle\ll L_{\rm event}$. The ratio of both length scales provides a  measure of the fraction of protons that can be annihilated. 

As seen in \rf{fig:proton-density}, from central to semi-peripheral ($N_{\rm part}\simeq 85$) collisions, the ratio of both lengths  nearly doubles. This means that  the annihilation fraction drops in semi-peripheral collisions to about half of the most central value. However,  the measured ratio p$/\pi$ is nearly constant over this range, increasing from $0.046\pm 0.003$ to $0.050\pm 0.003$. We interpret this as  experimental evidence that the net effect of $\mathrm{p\overline{p}}$  formation and annihilation is insignificant.  Therefore,  the annihilation of $\mathrm{p \overline{p}}$ pairs cannot serve as the explanation of the disagreement between the equilibrium {\small SHM} and observed small value of  p$/\pi$ ratio.  

Our estimate of the annihilation effect based on \req{annih-fraction} and the result seen in \rf{fig:proton-density}  is consistent with  the  annihilation effect reported in Ref.~\cite{Karpenko:2012yf}, where  detailed balance reactions forming  $\mathrm{p\overline{p}}$  were not considered. In this work, p$/\pi$ rises to p$/\pi=0.058$ already in the 20--30\% centrality bin ($N_{\rm part}=185$), which is more than 3 s.d. above experimental data (see \rf{fig:pOverPi}a).
Another work Ref.~\cite{Pan:2012ne} addresses directly our scenario of describing the experimental p$/\pi$ ratio and shows that with annihilation the required temperature would be  $T=165\pm5\,\mathrm{MeV}$ while without baryon annihilation a hadronization temperature of $T=145\pm5\,\mathrm{MeV}$ is required (initial yield  from equilibrium SHM). Such models of post-hadronization interactions also predict depletion of $\Xi$ yield and enhancement of $\Omega$ yield~\cite{Becattini:2012xb,Steinheimer:2012rd,Pan:2012ne}, which leads to even greater discrepancy between at least one of the multistrange baryons and equilibrium {\small SHM} predictions, since these yields as obtained before annihilation are already in general below the experimental data (see \rf{fig:firstfit}a,c). 

We do not see a scenario that would allow equilibrium SHM with hadronic afterburners to remain a viable model which can explain   a) the reduction of p$/\pi$ ratio from equilibrium {\small SHM} value as a function of centrality, and b) the yields of the multistrange baryons at the same time. On the other hand, the experimental value of p$/\pi$ ratio was predicted~\cite{Rafelski:2010cw}. The experimental result, the almost centrality independent p$/\pi$ ratio seen in figure \ref{fig:pOverPi} (note that the scale is greatly enhanced)  is now successfully fitted within non-equilibrium SHM in this work without any modifications to the model or essential change in model parameter values.

\subsection{\label{sec:physicalproperties}Fireball bulk  properties}
In order to obtain the bulk physical properties of the source of hadronic particles, we use exactly the same set of particles and the same assumptions about their properties as we employed in the fit procedure. Therefore, the physical properties we determine are consistent with the particle yields that originated our fit.  In other words, we sum the energy, entropy, etc. carried away by the observed particles,  adding to this observed yield the contributions due to  unobserved particles used in the {\small SHM} fit.

\begin{figure}[!bt]
\includegraphics[width=\columnwidth]{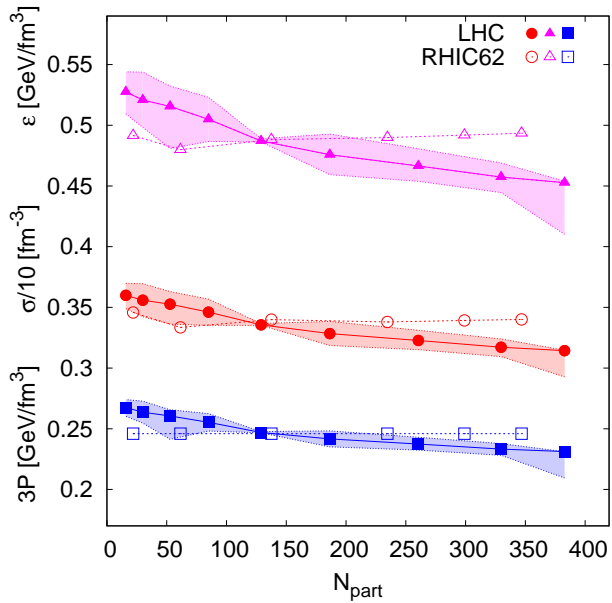}
\caption{\label{fig:physicalproperties}(color online) Bulk properties of the fireball as a function of centrality: from top to bottom,  energy density $\varepsilon$ (purple triangles), the entropy density $\sigma$ (red) circles, scaled down by factor 10, and the hadronization pressure $3P$ (blue) squares,  {\small LHC}2760 values shown with full symbols, {\small RHIC}62 with open symbols for comparison. Shaded areas show our estimate of systematic error arising from the uncertainty of~$\gamma_s$.}
\end{figure}

The bulk physical properties of the hadronizing fireball, that is energy, pressure, entropy and strangeness per entropy content are shown in the bottom part of  table~\ref{tab:params1} and in \rf{fig:physicalproperties} where shaded domains show our error estimate. Solid symbols are results of the fit, lines guide the eye.  In our {\small SHARE}v2 fit with {\small MINOS} minimization, the largest  uncertainty seen in table \ref{tab:params1} is the $\gamma_s$ and  $dV/dy$ error, see \rf{fig:thermalparameters}, other statistical bulk properties have relatively insignificant  errors. As can be seen in table~\ref{tab:params1}, multi-dimensional fits to data can result in nearly all of the fit error accumulating in the uncertainty of two or even just one parameter. In our fits, we see that the  dominant uncertainty is in the volume normalization. 

When error is  found in a few if not only one parameter, we checked for  uncertainty  arising within an  experimental data stability test. We test how a fit is modified when a small subset of experimental data points is altered arbitrarily but within error. We find that fits  comprising input data with such arbitrary modification  have in general larger errors distributed among all parameters. The convergence of the intensive parameters (e.g., $T$) in our initial fit suggests only a very small statistical error inherent to the data, while the extensive parameters (e.g., $V$) show a large error common to  particle yield  normalization.  In this situation, predicted ratios of hadron species should be more precise than their individual errors suggest. This is due to the experimental normalization of particle yields being, as this study indicates, strongly correlated. The presence of not vanishingly small error in $\gamma_s$ could be a signal of additional source of strange hadrons, for example charm hadron decays.

All fit errors propagate into the properties seen in \rf{fig:physicalproperties}. Since in~\rf{fig:thermalparameters} we consider densities, the error in volume does not affect these values.  Therefore, by recomputing the properties of the fireball shifting alone the value of $\gamma_s$ within one s.d., we obtain a good error evaluation  in the measurement of the bulk physical properties shown  in~\rf{fig:physicalproperties}.  The   point that stands out with very small error is at $N_{\rm part}=130$. This anomaly is due to accidental appearance of a sharp minimum in the highly non-trivial 7-dimensional parameter space. 

We are interested in studying the bulk properties of the source of hadrons in order to test the hypothesis that a {\small QGP} fireball was the source of particles observed. For this to be true, we must find appropriate   magnitude of bulk properties consistent with lattice results, and at the same time, a variation as a function of centrality that makes good sense. We observe in~\rf{fig:physicalproperties}  a smooth and slow decrease of energy density $\varepsilon$ (top), entropy density $\sigma$ (middle) and hadronization particle pressure $P$ (bottom) as a function of centrality. This slow systematic decrease of all three quantities is noted in particular comparing to {\small RHIC}62 (open symbols), where the properties seem to vary less. This maybe interpreted as an effect of  volume expansion at {\small LHC} leading to larger supercooling for larger systems.

The local thermal energy density of the bulk is the source of all particles  excluding the expansion flow kinetic energy. The value we find is $\varepsilon\simeq 0.50\pm0.05\,\mathrm{GeV/fm^3}$ in the entire centrality range. Nearly the same value is found within the chemical non-equilibrium approach for {\small RHIC}62~\cite{Petran:2011aa} and {\small RHIC}200~\cite{Rafelski:2004dp}.  We note  that $\varepsilon$  assumes the smallest value for the most central collisions, see table~\ref{tab:params1} and \rf{fig:physicalproperties}.  The hadronization pressure $P$ and entropy density $\sigma$ are also decreasing for more central collisions, which is consistent with our reaction picture of expanding and supercooling fireball --- the larger system in central collisions exhibits more supercooling reflected by a decrease of hadronization temperature and the above mentioned behavior of bulk properties. The error band is (as for $\varepsilon$) based on $\gamma_s$ uncertainty.

In the last row of table~\ref{tab:params1}, we show that entropy yield at {\small LHC}2760 is more than 3 times greater than obtained at {\small RHIC}62. The entropy yield $dS/dy$ as a function of participant number is shown in \rf{fig:dSdy}, and the notable feature is that the power law parametrization displays a nearly linear dependence at {\small RHIC}62 while at {\small LHC}2760 a strong additional entropy yield,  associated with the faster than linear increase,  is seen: $dS/dy\propto N_{\rm part}^{1.184}$. Most of the entropy is produced in an initial state mechanism which remains to be understood and our finding of the nonlinear entropy growth with $N_{\rm part}$ adds to the entropy production riddle and important observational result. 

\begin{figure}
\includegraphics[width=\columnwidth]{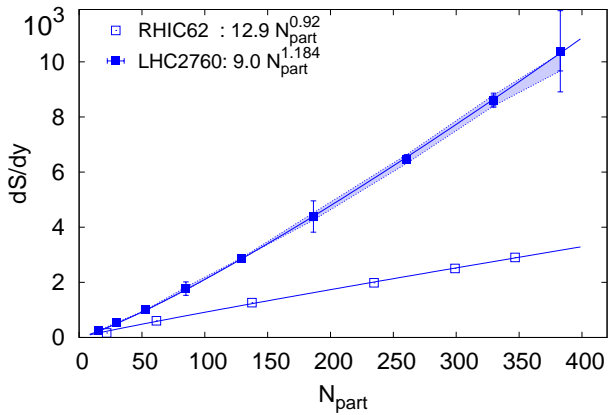}
\caption{\label{fig:dSdy}(color online) Entropy yield $dS/dy$ at  {\small LHC}2760 and at {\small RHIC}62  as a function of centrality  showing power law fit parameters in the legend. Colored band represent uncertainty based on $\gamma_s$ fit uncertainty. Error bars arise from error in the volume $dV/dy$. }
\end{figure}

However,   at {\small LHC}2760, one expects a component in the entropy count arising from the inclusion  of the decay products of heavy charmed hadrons in the hadron yield. This entropy component is different from entropy produced in initial reactions,  this is the entropy arising from hard parton collision  production of charm, and  post-hadronization decay of charmed hadrons. It is unlikely  that the non-linearity of the entropy yield is due to this phenomenon as one can easily see that the required charm yield would be very large. We will return, in near future, to this question.  The uncertainty of entropy depicted in \rf{fig:dSdy} as a shaded band is based  alone  on $\gamma_s$ variation, as was obtained for other  physical properties in \rf{fig:physicalproperties}. A further   error due to variance in $dV/dy$ is shown as a separate error bar. Where it is invisible for the  {\small LHC}2760, it is hidden in symbol size.

We turn now  to study strangeness per entropy $s/S\equiv (ds/dy)/(dS/dy)$ in the source fireball. We are interested in this quantity since both entropy and strangeness yields are preserved in the hadronization process. Therefore, by measuring $s/S$, we measure the ratio of strange quark abundance to total quark and gluon abundance which determines the source entropy, with a well known proportionality factor. For the presently accepted small strange quark mass $m_s(\mu=2{\rm GeV})=95\pm5$\,MeV~\cite{Beringer:1900zz}, the predicted value shown in figure 5 of Ref.~\cite{Kuznetsova:2006bh} is $s/S\simeq 0.0305\pm0.0005$. Finding this result in our {\small LHC} data analysis is necessary in order to maintain the claim that the source of hadrons is a rapidly disintegrating chemically equilibrated {\small QGP} fireball.

In the \rf{fig:sOverS}a, we show the strangeness per entropy $s/S$ in the source fireball. The solid squares are for the {\small LHC}2760, and open symbols for {\small RHIC}62. We see that $s/S$ saturates at $s/S\simeq 0.030$ at {\small LHC}2760, a value reached already for $N_{\rm part}> 150$, thus for a smaller number of participating nucleons than we found at {\small RHIC}62, and which value remains constant up to the maximum available $N_{\rm part}$.  This agrees with equilibrated {\small QGP} hypothesis and suggests that the source of hadrons was in the same conditions for a wide range of centrality. 

\begin{figure}[!tb]
\includegraphics[width=\columnwidth]{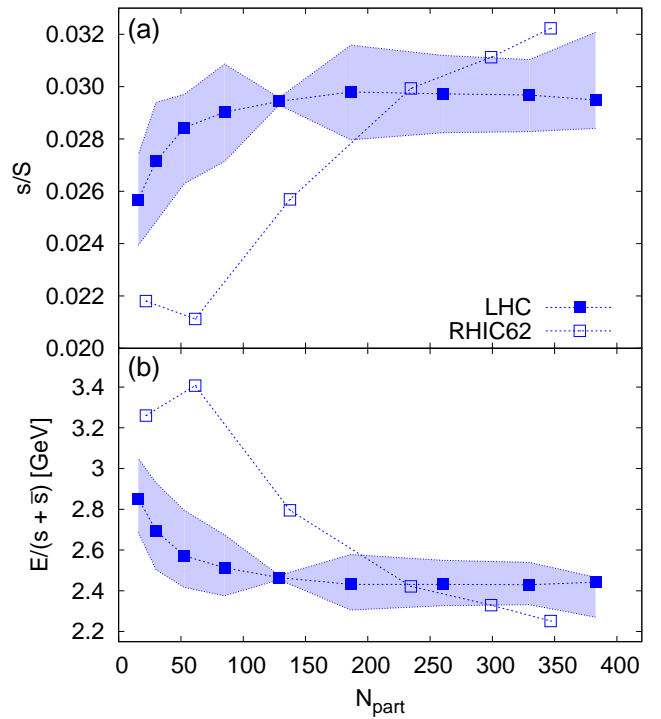}
\caption{\label{fig:sOverS}(color online) Panel (a): strangeness per entropy $s/S$ content of the fireball at {\small LHC}2760 (filled squares) and at {\small RHIC}62 (open squares) as a function of centrality; panel (b): the thermal energy cost to make a strange--anti-strange quark pair. Colored bands represent uncertainty based on $\gamma_s$ uncertainty.}
\end{figure}

This  constant $s/S$ value as a function of centrality  can be interpreted as an evidence of chemical equilibrium for a {\small QGP} source: the strangeness yield normalized to all quark and gluon yield inherent in $S$ can be constant only if dynamical processes find a chemical balance for the differently sized fireballs. The  value $s/S=0.03$ is in excellent quantitative agreement with microscopic model of strangeness production and equilibration in {\small QGP}~\cite{Kuznetsova:2006bh,Letessier:2006wn}, adopting latest  strange quark mass value. The  high {\small QGP} strangeness yield oversupplies in hadronization   the hadron phase space resulting in $\gamma_s\simeq 2$ seen in \rf{fig:thermalparameters}.   Considering the   {\small RHIC} results shown in \rf{fig:sOverS}a, we see  a slightly  higher  $s/S$ saturation limit for most central collisions, though the difference is within the {\small RHIC} error band (not shown). It is possible that  $s/S$ {\small LHC}2760 result is 5--10\%  diluted due to inadvertent inclusion in the entropy count of the charm decay hadrons. It is also of interest to note that  at {\small RHIC}62, $s/S$ increases monotonically (discounting  the low confidence level most peripheral point) with increasing $N_{\rm part}$ suggesting that the {\small QGP} source reaches chemical equilibrium only for most central collisions. At  {\small LHC}2760 there is such increase  for much lower size volume of the collision centrality  $N_{\rm part}< 150$.

In~\rf{fig:sOverS}b, we show the  thermal energy cost to make a strange quark--anti-quark pair. At {\small LHC}2760, the energy cost to make a strange pair is practically constant for  the wide range of mid-central to central collisions, which confirms  that strangeness  in the {\small QGP} fireball is in chemical equilibrium at the time of hadronization. The slight increase of the thermal energy cost for small centralities corresponds to the lower yield of strangeness seen in \rf{fig:sOverS}a. At {\small RHIC}62, we see monotonically improving energy efficiency converging to a value slightly below our new {\small LHC}2760 result, but well within the error bar at {\small RHIC}62 (not shown). The rise of energy cost for  smaller systems relates to the fact that a larger  and notable fraction of strangeness was  produced in first hard collision processes during the initial stages of the collision which for {\small RHIC}62 and {\small LHC}2760  is resulting in higher energy needed to produce one strange--anti-strange pair.

\subsection{\label{sec:lattice}Connection to lattice results and related considerations}
Elaborate lattice-{\small QCD} numerical computations of {\small QGP}--hadron transition regime  are available today~\cite{Endrodi:2011gv,Bazavov:2011nk}, and are comprehensively reviewed in Ref.~\cite{Philipsen:2012nu}: the Hot{\small QCD} collaboration~\cite{Bazavov:2011nk} converged for 2+1 flavors towards $T_c=154\pm9$ MeV. The question how low the value of $T_c$ can be, remains in current intense discussion, as the latest work of Wuppertal--Budapest collaboration~\cite{Borsanyi:2012rr} suggests a low $T_c\simeq  145 $\,MeV. For an expanding {\small QGP} with supercooling, this can lead to hadronization below  $T_c\simeq  145 $\,MeV and near $T=140$ MeV. This is indeed the range of values of $T$ that we find in our chemical non-equilibrium {\small SHM} analysis. 

A comparison of lattice results with freeze-out conditions is shown in~\rf{fig:phasediagram}.  The two bands near to the temperature axis display the lattice critical temperature in the range $T_c=154\pm9\,\mathrm{MeV}$~\cite{Bazavov:2011nk} (red online) and  $T_c=147\pm5\,\mathrm{MeV}$~\cite{Borsanyi:2012rr} (green online). The symbols show the results of hadronization analysis in the $T$--$\mu_B$ plane. We selected here the results for the most central collisions and heaviest nuclei.  The  solid (blue) circles are {\small SHARE} chemical non-equilibrium results obtained by our group, with result presented in this paper included in the {\small LHC} domain, and {\small RHIC} and {\small SPS}  results seen, e.g., in~\cite{Letessier:2005qe,Rafelski:2009jr,Rafelski:2009gu}. The {\small LHC}2760 freeze-out temperature is in our case clearly below the   lattice critical temperature $T_c$. As just discussed, this is expected for supercooling followed by sudden hadronization. We show also $\gamma_q=1$ results of other groups: {\small GSI}~\cite{BraunMunzinger:1994xr,Andronic:2005yp}, Florence~\cite{Becattini:2010sk,Manninen:2008mg,Becattini:2003wp}, {\small THERMUS}~\cite{Cleymans:2005xv}, {\small STAR}~\cite{Abelev:2008ab} and {\small ALICE}~\cite{Abelev:2012wca}.  These results show the chemical freeze-out temperature in numerous cases well above the lattice critical temperature $T_c$, which in essence means that these {\small SHM} calculations are incompatible with lattice calculations.

\begin{figure}[!bt]
\includegraphics[width=\columnwidth]{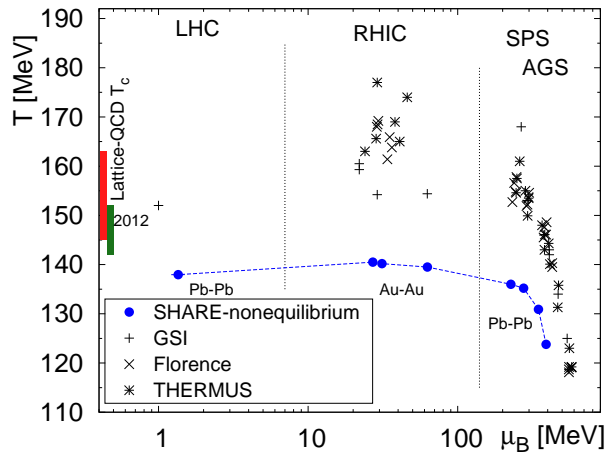}
\caption{\label{fig:phasediagram}(color online) Phase diagram showing current lattice value of critical temperature $T_c$ calculated by two groups~\cite{Borsanyi:2012rr,Bazavov:2011nk}, and results of this work as well as our previous results (blue circles)~\cite{Rafelski:2009jr,Rafelski:2009gu,Letessier:1998sz} and results of other groups~\cite{BraunMunzinger:1994xr,Andronic:2005yp,Becattini:2010sk,Becattini:2003wp,Manninen:2008mg,Abelev:2012wca,Cleymans:2005xv,Abelev:2008ab}. Full circles refer to chemical non-equilibrium, all other symbols refer to fit results with chemical equilibrium of light quarks.}
\end{figure}

The two recent lattice results, shown in~\rf{fig:phasediagram}, challenge  the chemical equilibrium hadronization~\cite{BraunMunzinger:2001ip} scenario widely used for the past decade, which produces a hadronization temperature above the lattice phase cross-over results. Two conspiring  hypotheses were made in Ref.~\cite{BraunMunzinger:2001ip}: 1) there is  chemical equilibrium  in; 2) a long lived hadron gas phase. Both statements were assumptions without  theoretical or experimental evidence `confirmed' by fits to data, which had even with the large experimental errors a rather large $\chi^2$ and thus a negligible confidence level. Therefore, this model needed additional support. Lattice results showing $T_c=173\pm 8$ MeV were often introduced in support of equilibrium-{\small SHM}. Such a high $T_c$ appears, for example, in figure 10 in Ref.~\cite{de Forcrand:2002ci}, but reading the text, one sees that it applies to the mathematical case of two light quark flavors on discrete space-time.  Allowing for strangeness flavor in {\small QGP}, the hadronization temperature must decrease. Therefore,  already a decade ago  $T_c=154\pm 8$ MeV was the best estimate for 2+1 flavors, leading  to the consensus range $T_c=163\pm 15$~MeV before continuum limit. Present day continuum value we estimate to be $T_c\simeq 150\pm7$\,MeV combining the two results seen in~\rf{fig:phasediagram}.

An important requirement, for the full chemical non-equilibrium hadronization approach, is that in the hadronization process,  quark flavor abundances emerge as produced at an earlier and independent stage of fireball evolution. Our analysis relies on  hadronization being fast, not allowing a significant modification of the available quark abundances.  These quark abundances at {\small LHC} in a wide range of centralities and in most central  {\small RHIC}  collisions are near to the {\small QGP} chemical equilibrium abundance. In order for the quark yields to remain largely unchanged during  hadronization and after,  it is necessary that the transformation from {\small QGP} to hadrons (hadronization) occurs  suddenly and at a relatively low temperature, near to the expected chemical freeze-out point where particle abundances stop evolving. The two pion correlation experimental results favor sudden hadronization, which has been seen in the results for a long time~\cite{Csernai:2002sj}. The sudden hadronization model was required for consistency with these results~\cite{Csernai:1995zn,Rafelski:2000by}. It is associated with  chemical non-equilibrium  {\small SHM} analysis of the data~\cite{Letessier:1998sz,Letessier:2000ay}. Today, with lattice {\small QCD} transition conditions reaching a low $T$ consensus, the only {\small SHM} approach that remains valid is the chemical non-equilibrium.

\section{\label{sec:conclusion}Discussion}
\subsection{What is new at LHC}
The primary difference between {\small RHIC}62 and {\small LHC}2760 data is a $4$-times larger transverse volume $dV/dy$ at hadronization, as seen in \rf{fig:thermalparameters}a. Increase of volume at {\small LHC} compared to {\small RHIC}, rather than a change of hadronization temperature, shows a common source of hadrons, a signature of {\small QGP} formation.
The increased volume is in qualitative agreement with the two pion correlation studies~\cite{Aamodt:2011mr}. Given the nearly constant entropy density at hadronization, the growth of volume drives the total entropy yield, which is up to $3.2$ times greater at {\small LHC}2760 than at {\small RHIC}62. 

Other differences of {\small LHC}2760 compared to {\small RHIC}62 are: 
1) An order of magnitude  smaller baryochemical potential $\mu_{\rm B}\simeq 1.5\,$MeV, see \rf{fig:muBmuS}.\\
2) Phase-space occupancy $\gamma_q$ constant as a function of centrality.\\
3) Earlier saturation of $\gamma_s$ as a function of centrality, and thus $\gamma_s/\gamma_q$-ratio following the behavior of $\gamma_s$.\\ 
For comparison, at {\small RHIC}62, we have a fast increase of $\gamma_s$ over the entire range of $N_{\rm part}$, as is shown in \rf{fig:thermalparameters}c. The {\small LHC}2760 result is interpreted to mean that  the {\small QGP} fireball is rapidly chemically equilibrated already for small $N_{\rm part}$, while at {\small RHIC}62, we must have a large value of $N_{\rm part}$, that is a large volume, and thus large lifespan, to achieve full strangeness chemical equilibrium in the {\small QGP} fireball. The  value $s/S=0.03$ is in excellent qualitative agreement with microscopic model of strangeness production and equilibration in {\small QGP} and the associated predictions of the final state yield~\cite{Kuznetsova:2006bh,Letessier:2006wn}. 

As a comparison of our present work with our predictions~\cite{Rafelski:2010cw} shows, the yield of strangeness is $\sim 20\%$ below our prior expectations. These were motivated by consideration of a very rapidly diluting  {\small QGP} fireball, wherein the early strangeness {\small QGP} equilibrium is preserved and leads to overabundance, above {\small QGP} chemical equilibrium at time of hadronization. Such behavior was indicated given the {\small RHIC} results showing a steady rise, see~\rf{fig:sOverS}a for {\small RHIC62}. Instead, we find a perfectly equilibrated {\small QGP} fireball:   the observed value of $s/S\simeq 0.03$ is expected for a  chemically equilibrated QGP fireball near hadronization condition. This equilibrium {\small QGP} saturated value $s/S=0.03$ is observed for many centralities. Since to obtain our prediction we used  $s/S=0.037$, both the value of $\gamma_s$ and yields of Kaons are equally $\sim 20\%$ suppressed compared to expectation~\cite{Rafelski:2010cw}, and  other strange particles as well. How this is possible will be one of the riddles that future data and theoretical modeling will need to address. For us, this strangeness suppression compared to expectation  is the most remarkable difference from {\small RHIC} data that we have found in this first {\small LHC} result analysis.

\subsection{Centrality dependence}
Considering the bulk properties of the fireball at hadronization, the most remarkable finding is that there is so little centrality dependence. This means that at {\small LHC}2760 the source of hadrons is a hot drop of energy that varies mainly in volume as we vary the collision geometry. This applies to energy density   $\varepsilon\simeq 0.50\pm0.05\,\mathrm{GeV/fm^3}$, hadronization pressure $P$ and entropy density $\sigma$ in the entire centrality range, see table~\ref{tab:params1} and \rf{fig:physicalproperties}.  These bulk properties decrease monotonically and slowly and assume  the smallest value for the most central collisions, supporting the reaction picture of expanding and supercooling fireball --- the larger system supercools a bit more. Recall that the error bands  in~\rf{fig:physicalproperties} are based on $\gamma_s$ uncertainty. The one  clear centrality dependence of the fireball we find is the rapid rise and early appearance of the strangeness yield saturation seen in~\rf{fig:sOverS}a.

The chemical freeze-out temperature $T$ decreases by about 3\,$\MeV$ at all centralities compared to {\small RHIC}62, see middle panel in \rf{fig:thermalparameters} (we do not consider here the most peripheral {\small RHIC}62 result which has small confidence level). We believe that this result is related to the need to expand and supercool further the initial energy and entropy rich {\small LHC}2760 fireball. The large expanding {\small QGP} matter pushes further out, supercooling more and yielding a further reduction in the sudden hadronization temperature. The freeze-out temperature $T$ increases towards more peripheral collisions, see \rf{fig:thermalparameters}b, which can be explained by the disappearance of   supercooling present for the  most central and most energetic collision systems. Considering the behavior of both {\small LHC}2760 and {\small RHIC}62 for $N_{\rm part}\to 0$, we obtain $T_{\mathrm{had}} \to 145\pm4\MeV$, applicable to hadronization without supercooling. This value is in good agreement with the latest lattice result~\cite{Borsanyi:2012rr} for transformation temperature from {\small QGP} to hadrons.

The value of temperature and its behavior as a function of centrality and heavy ion collision energy suggest that produced hadrons emerge directly from a sudden break-up of quark--gluon plasma. The hadron particle density at this low $T$ is sufficiently low to limit the particle number changing reactions and render these insignificant. $T=145$--$140$ MeV is at, and below, the expected {\small QGP} phase transition. The presence of chemical non-equilibrium at this low $T$ means that hadrons did not evolve into this condition, but must have been produced directly from the deconfined phase. This is consistent with the two pion correlation time-parameter, which suggests that particles are produced at a scale which is sudden compared to the size of the system, as is expected for a supercooled {\small QGP} state undergoing, e.g., a filamenting breakup at $T\simeq 140$~MeV, and the result of such dynamics is qualitatively consistent with the features described here~\cite{Csernai:2002sj}. 

The second to last row in table~\ref{tab:params1} shows the ratio of entropy at {\small LHC}2760 to {\small RHIC}62, $S_{\mathrm{LHC}}/S_{\mathrm{RHIC}}$, within the rapidity interval $-0.5\le y\le 0.5$. The entropy enhancement factor increases monotonically with centrality, from ratio of 1.27 in the most peripheral bin to ratio 3.23 in the most central bin.
This increase requires volume dependent additional entropy production mechanisms, which are more effective for the more central, larger $N_{\rm part}$, collisions. Such an  increase can arise from hard parton collision generated jets, which are better quenched in the larger volume of matter, and in addition in abundant charm production, which decays into hadrons and appears  as additional hadron multiplicity, i.e., entropy. As long as the additional entropy is generated in early stages of the fireball evolution, this has little impact on {\small SHM}  method of approach in study of hadronization. For example, the quenching of {\small QCD} jets feeds thermal degrees of freedom that can convert a part of its energy into strangeness. However, charm decay is different as it occurs after hadronization. Thus, it needs to be accounted for and/or proved irrelevant. It is possible that charm decay entropy generating mechanism may be the cause of the slight (5\%) strangeness $s$ over entropy $S$ dilution at {\small LHC}2760 (see \rf{fig:sOverS}a).

\subsection{What we learn about hadronization at LHC}
The full chemical non-equilibrium  is introduced by the way of the  parameter $\gamma_q\ne 1$. This allows one to describe  a situation in which a source of hadrons disintegrates faster than the time necessary to re-equilibrate the yield of light quarks present.   The two pion correlation data provide  experimental evidence that favors a rapid breakup of {\small QGP} with a short time of hadron production~\cite{Csernai:2002sj}, and thus favors very fast, or sudden, hadronization~\cite{Csernai:1995zn,Rafelski:2000by}.  There has been for more than a decade an animated discussion if the parameter $\gamma_q$ is actually needed with arguments such as simplicity used to invalidate the full chemical non-equilibrium approach.

We have shown that {\em only} the chemical non-equilibrium {\small SHM} describes very well all available {\small LHC}2760 hadron production data obtained in a wide range of centralities obtained  in the   rapidity interval $-0.5\le 0\le 0.5$, and the outcome is consistent with lattice {\small QCD} results. We successfully fit the data with $\chi^2/\mathrm{ndf} < 1$ for all centrality bins, and show a smooth systematic behavior as a function of centrality of both, the statistical  {\small SHM}  parameters, see ~\rf{fig:thermalparameters}, and bulk physical properties, see~\rf{fig:physicalproperties}, that allow a simple and consistent interpretation. {\small SHM} is validated at  {\small LHC}2760 as it describes precisely yields of different particles in a wide range of collision centrality and which span over more than 5 orders of magnitude, see \rf{fig:fittedparticles}.

We have shown that it is impossible to fit the ratio p$/\pi = 0.046\pm0.003$~\cite{Abelev:2012vx,Abelev:2012wca} together with the other data, when choosing a {\small SHM} with $\gamma_q=1$. However, p$/\pi \simeq 0.05$  is a natural outcome of our chemical non-equilibrium fit where $\gamma_q\simeq 1.6$.  This result was predicted~\cite{Rafelski:2010cw}: within the  chemical non-equilibrium {\small SHM}, p$/\pi|_{\rm prediction} = 0.047\pm 0.002$ for $P=82\pm 5$\,MeV/fm$^3$ is in agreement with experimental result we discuss here, for most central collisions p$/\pi|_{\rm\scriptsize ALICE} = 0.046\pm0.003$.

We have discussed, in section~\ref{sec:pOverPi}, the possibility  of p$/\pi$ ratio evolving after hadronization, and found this scenario to be highly  unlikely considering that experimental ratio p$/\pi$ does not vary in a wide centrality domain. Therefore, the fact that chemical equilibrium {\small SHM} variant over-predicts p$/\pi$ {\em and} produces a poor $\chi^2_{\rm total}$, see \rf{fig:pOverPi}b, demonstrates that the chemical equilibrium {\small SHM} approach (with or without post-hadronization interactions) does not work at {\small LHC}2760. Further  evidence for the chemical non-equilibrium {\small SHM} comes from universality of hadronization at  {\small LHC}2760 and at {\small RHIC}, see subsection \ref{sec:physicalproperties} and Ref.\cite{Petran:2013qla}.

\subsection{Predicting experimental results}\label{predict_method}
Our prediction of hadron yields~\cite{Rafelski:2010cw}  required as input the charge particle multiplicity $dN_{\rm ch}/dy$ which normalizes the reaction volume $dV/dy$. Further, we assumed strangeness per entropy content $s/S$, and the nearly universal hadronization pressure with preferred value $P = 82\pm 5$\,MeV/fm$^3$. This is  accompanied by the strangeness conservation constraint $\langle s-\bar s\rangle=0$ and  the projectile--target charge to baryon ratio $Q/B=0.4$ and, as baryochemical potential cannot yet be fully defined, an approximate value $\cal O$(1)\,MeV. Using this input with a 5\% error, we obtain the most compatible  values of $dV/dy,\,T,\,\gamma_q,\, \gamma_s$ and chemical potentials, and we can evaluate the particle yields along with  fireball properties. 

We have redone the predictions for $\sqrt{s_{NN}}=2.76\,\mathrm{TeV}$ case with the tested and released {\small SHARE}v2.2 code and find that the pre-release {\small SHARE}  predictions in~\cite{Rafelski:2010cw} were made for $dN_{\rm ch}/dy=2150$ and not for $dN_{\rm ch}/dy=1800$. Therefore, all absolute hadron yields stated in Ref.~\cite{Rafelski:2010cw} are normalized to be $\sim\!20\%$ too large, in addition to the strangeness overcount originating in the assumption $s/S=0.037>0.030$. The  ratios of hadrons with the same strangeness content were correctly predicted.

Applying our prediction method using the updated strangeness value of $s/S=0.030$ and a more precise hadronization pressure estimate $P\simeq 77\pm 4,\mathrm{MeV/fm}^3$ results, for $\sqrt{s_{NN}}=2.76\,\mathrm{TeV}$,  in accurate prediction of all  hadron particle yields, statistical parameters, and fireball bulk properties, without using as input any individual hadron yield. This validates our approach~\cite{Rafelski:2010cw}, which can be applied to the forthcoming Pb--Pb collisions at $\sqrt{s_{NN}}=5.5\,\mathrm{TeV}$ or in the {\small RHIC} beam energy scan.  Noting that the multiplicity of produced hadrons is synonymous to entropy of the fireball, this result means that all hadron yields can be predicted within the framework of chemical non-equilibrium {\small SHM} using as input the  properties of the bulk matter in the fireball.

\subsection{Conclusions and outlook}

We have shown that the non-equilibrium SHM model in the {\small LHC}  reaction energy range is yielding a very attractive data fit. We have argued that non-equilibrium SHM  is today  favored by the lattice results, since  we must have $T<T_c$,  and  lattice is moving lower in $T_c$, see $T_c=147\pm5\,\mathrm{MeV}$~\cite{Borsanyi:2012rr}.  Only the non-equilibrium SHM range  $T<145\,\mathrm{MeV}$ remains convincingly  compatible.  Considering the dynamics of the fireball expansion  $\Delta T\equiv T_c-T$ is of magnitude   where we would like it for supercooling. Moreover, the chemical non-equilibrium {\small SHM} is favored by offering  simplicity, as it needs no after-burners. Ockham's razor argument (lex parsimoniae) can be used to conclude that non-equilibrium SHM is a valid precise description of multi-hadron production.

The good fit within the realm of non-equilibrium SHM of all observed particles allows us to predict with some confidence the yields of yet unmeasured hadrons within the chemical non-equilibrium {\small SHM} scheme, which are seen in table~\ref{tab:yields1}. The question is how stable these yields are when data basis of the fit increases to include new measurement. A small {\small SHM} parameter change should be expected  also when we  refine the theoretical model by adding features, such as inclusion of hadrons from perturbative {\small QCD} jets and/or charm hadron decay contribution to hadron yields. We believe that predictions for the primary `stable' hadrons such as $\eta$ are accurate. On the other hand, even the  minor changes in {\small SHM} parameters can  have relatively large effect especially for anti-matter clusters shown in the bottom part of table \ref{tab:yields1}:  in the anti-alpha, we have 12 anti-quarks, and a few \% error in understanding their primordial yield is raised to 12th power. 

It is quite remarkable that despite a change by a factor of 45 in reaction energy, we find for all centralities at both {\small LHC}2760 and {\small RHIC}62, that the energy density of hadronizing matter is $0.50\pm0.05$ GeV/fm$^3$, as is seen in~\rf{fig:physicalproperties}.  In fact, the present day data favor a systematic decrease of hadronization pressure $P$ from peripheral towards central collisions as compared to earlier  {\small RHIC}62~\cite{Petran:2011aa}, {\small RHIC}200~\cite{Rafelski:2004dp} and our preliminary {\small LHC} analysis with limited data set~\cite{Petran:2013qla}. It is possible that the more dynamical expansion of the {\small LHC}2760 fireball and deeper supercooling of the fireball are the cause. 

We checked that assuming universal hadronization pressure, we could obtain a very good fit to particle data for all centrality {\small LHC}2760 data bins. This means that if and when more hadron yield data are available, the decrease in bulk properties with centrality seen in \rf{fig:physicalproperties} could easily disappear. Therefore, the presence of a constant critical hadronization pressure~\cite{Rafelski:2009jr} could extend from {\small SPS} to  {\small LHC}.  We are investigating this hypothesis, as well as the possibility that another quantity governs universality of hadronization. We hope to return to the matter as soon as we  have understood better the final state contributions to hadron yields from charmed hadron decays.

We have shown that the precise hadron yields measured by the {\small ALICE} collaboration at {\small LHC}2760 have offered a vast new opportunity to explore the properties of the {\small QGP} fireball and to understand the dynamics of its evolution and matter production.  We are able to quantify the key physical properties at this early stage. With more data becoming available, we expect a significant refinement and improved understanding of both the {\small QGP} fireball and mechanisms of matter creation out of the deconfined {\small QGP} phase.


\subsection{Update} We have checked that the new results~\cite{Abelev:2013xaa,ABELEV:2013zaa} on strange hadron multiplicities which became available at the beginning of the SQM2013 meeting end of July 2013 are fully compatible:  the K$_S$, $\Omega$, and $\Xi$  are in remarkable agreement with our here presented evaluations and $\Lambda$ yield is as much off as is our fitted preliminary $\Lambda/\pi$ ratio, see \rf{fig:KstarLambdaRatios}, that is the theoretical  $\Lambda$ yield is in general about 1.2 s.d.  smaller compared to the   final experimental $\Lambda$ yield. Here we note that the presented fits are carried out without taking into account charmed particle decay products, which beyond the  generally enhanced overall hadron multiplicity  produce a non-negligible number of additional strange baryons.

\section*{\label{sec:acknowledgements}Acknowledgments}
This work has been supported by a grant from the U.S. Department of Energy, DE-FG02-04ER41318, Laboratoire de Physique Th{\' e}orique et Hautes Energies, LPTHE, at University Paris 6 is supported by CNRS as Unit{\' e} Mixte de Recherche, UMR7589. JR thanks CERN-PH-TH for the hospitality while most of this work was carried out. JR thanks members of {\small ALICE} collaboration, in particular Karel Safarik, Boris Hippolyte and Federico Antinori, for many conversations and clarifications.

\begin{table*}[!hbt]
\caption{\label{tab:experimental} Enhancement of multi-strange baryon yields per participant pair relative to p--p collisions, p--p yields and calculated yields in Pb--Pb, which we use as input to our interpolation as a function of $N_{\rm part}$.} 
\begin{ruledtabular}
\resizebox{\textwidth}{!}{ 
\centering
\begin{tabular}{ l | l | c | c | c | c | c }
Particle & Ref. & Centr. & $\langle N_{\rm part}\rangle$ & p--p data $(dN/dy)_{\mathrm{pp}}$ and enhancement $E$ & \multicolumn{2}{c}{$(dN/dy)_{\mathrm{PbPb}} = E\,0.8\,(dN/dy)_{\mathrm{pp}} (\langle N_{\rm part}\rangle/2) $} \\ \hline
\multirow{5}{*}{$\Xi^-$}		&  \cite{Abelev:2012jp} & p--p        & 2    & $(8.0 \pm 0.7)\times 10^{-3}$ \\ \cline{2-5}
								&	\multirow{4}{*}{}  & 60--90\% & 17.6 & $E=1.58\pm 0.18$ & \multicolumn{2}{c}{$0.090\pm0.010$} \\
                               &                                             & 40--60\% & 68.8 & $E=2.48\pm 0.26$ & \multicolumn{2}{c}{$0.55\pm0.06$}\\
                               &   \cite{Chinellato:2012jj}                                           & 20--40\% & 157  & $E=2.95\pm 0.32$ & \multicolumn{2}{c}{$1.51\pm0.17$} \\
                               &                                              &  0--20\% & 308  & $E=3.08\pm 0.33$ & \multicolumn{2}{c}{$3.08\pm0.33$}\\    \hline
\multirow{5}{*}{$\Xibar^+$}   &      \cite{Abelev:2012jp}      & p--p        & 2    & $(7.8 \pm 0.7)\times		 10^{-3}$  \\ \cline{2-5}
								&	\multirow{4}{*}{}  & 60--90\% & 17.6 & $E=1.57\pm 0.19$ & \multicolumn{2}{c}{$0.087\pm0.011$} \\
                               &      & 40--60\% & 68.8 & $E=2.56\pm 0.26$ & \multicolumn{2}{c}{$0.56\pm0.06$} \\
                               & \cite{Chinellato:2012jj}                                             & 20--40\% & 157  & $E=3.20\pm 0.35$ & \multicolumn{2}{c}{$1.59\pm0.17$} \\
                               &                                              &  0--20\% & 308  & $E=3.00\pm 0.32$ & \multicolumn{2}{c}{$2.91\pm0.32$} \\ 
\hline
$\Omega^-$                     & \cite{Abelev:2012jp} \multirow{2}{*}{}       & \multirow{2}{*}{p--p}        & \multirow{2}{*}{2}   & $(0.67\pm 0.08)\times 10^{-3}$  \\
$\Omegabar^+$                  &                                              &         &     & $(0.68\pm 0.08)\times 10^{-3}$ & \hspace{1.5cm}$\Omega^-$\hspace{1.5cm} &  \hspace{1.5cm}$\Omegabar^+$\hspace{1.5cm} \\  \cline{1-5}
\multirow{4}{1.7cm}{ $$\frac{(\Omega^-\!\!+\Omegabar^+)}{2}$$ } &	\multirow{4}{*}{}  & 60--90\% & 17.6 & $E=2.56\pm 0.53$ & $0.012\pm0.003$ & $0.012\pm0.003$\\
                               & \cite{Chinellato:2012jj}                           & 40--60\% & 68.8 & $E=4.57\pm 0.79$ & $0.08\pm0.02$ & $0.08\pm0.02$ \\
                               &                                              & 20--40\% & 157  & $E=5.23\pm 0.95$ & $0.22\pm0.05$ & $0.21\pm0.04$ \\
                               &                                              &  0--20\% & 308  & $E=6.97\pm 1.27$ & $0.57\pm0.12$ & $0.56\pm0.12$ \\ 
\end{tabular}
} 
\end{ruledtabular}
\end{table*}

\appendix
\section{\label{sec:data}Data rebinning}
\subsection*{\label{ssec:yields}Rebinning multi strange hadron yields}
Since there is no literature stating explicitly the yields of $\Xi$ and $\Omega$ in Pb--Pb collisions, we proceed to obtain these results by unfolding the preliminary enhancement data. We combine the yield of $\Xi$ and $\Omega$ produced in p--p collisions at 7 TeV~\cite{Abelev:2012jp} stated in table~\ref{tab:experimental} and labeled `pp' in the third column therein, with the `preliminary' enhancement $E$ relative to p--p and normalized to a pair of participating nucleons shown in Ref.~\cite{Singha:2012qv} and which we also show in the fifth column of table~\ref{tab:experimental}. We generate the first data point for  the centrality bin 0--20\% by averaging the number of participants in the centrality bins from 0 to 20\% shown in table 1 of Ref.~\cite{Abelev:2013qoq}. We reduce the yields of both $\Xi$ and $\Omega$ by a constant factor of 0.8 in order 
to compensate for the difference in collision energy $\sqrt{s}=7\,\mathrm{TeV}$ in p--p collisions and $\sqrt{s_{NN}}=2.76\,\mathrm{TeV}$ in Pb--Pb.  We obtained the magnitude of this energy correction factor by comparing with the actual yield for the 0--20\% centrality bin given in Ref.~\cite{Ivanov:2013haa}. To disentangle the combined yield of $\Omega+\Omegabar$, we use the separated $\Omega$ and $\Omegabar$ yields from p--p collisions~\cite{Abelev:2012jp}, see   table~\ref{tab:experimental}.

We use the relative errors of the enhancements to estimate the errors of the multi strange baryon yields, that is $\sim\!11\%$ for $\Xi$ and $\sim\!20\%$ for $\Omega$. Our adopted $\Omega$ error is larger by $\sim\! 3\%$ than the error of its yield in 0--20\% centrality bin~\cite{Ivanov:2013haa}. We adopted this slightly increased error  to account for procedure which lead us to estimate the yield of  $\Omega$, $\Omegabar$ based in part on  $\Omega+\Omegabar$ yield.  The mathematical operations leading to the yields, the yields and widths we use are stated in self-explanatory fashion in table~\ref{tab:experimental}.

To account for the different centrality bins for   multi strange baryons as compared to $\pi$, K, p and $\phi/K$, we express the centrality bins in terms of average number of participants according to~\cite{Abelev:2013qoq} and then interpolate every particle yield $dN/dy$ available as a function of ${N_\mathrm{\rm part}}$ with a power law
\begin{equation}
\frac{dN\left({N_\mathrm{\rm part}}\right)}{dy} = a \,{N}_\mathrm{\rm part}^b + c,
\label{eq:interpolation}
\end{equation}
where $a,b$ and $c$ are free parameters. The form of the function has no immediate physical motivation, it  serves well the purpose of unifying the data across incompatible centrality bins. This method enables us to evaluate the invariant yields for any given ${N_\mathrm{\rm part}}$, {i.e.}, arbitrary centrality and thus enables us to include the multi-strange baryon yields in this analysis. Interpolation parameters together with $\chi^2$ of each particle interpolation are summarized in table~\ref{tab:interpolation}. For completeness, and potential future use, we present also the parametrization of $\pi^\pm$, K$^\pm$ and p$^\pm$ which do not require rebinning. Small values of $\chi^2$/ndf show that our description is accurate in the given interval of ${N_\mathrm{\rm part}}$.  Interpolation curves are depicted with dashed lines in \rf{fig:fittedparticles} for particle yields.  

\begin{table}[!bht]
\caption{\label{tab:interpolation} Interpolation parameters of particle yields as defined in \req{eq:interpolation}.} 
\begin{ruledtabular}
\resizebox{\columnwidth}{!}{ 
\centering
\begin{tabular}{l|c|c|c|c}
Particle & a & b & c & $\chi^2/{\rm ndf}$ \\
\hline
$\pi^+$  	 &   $0.725\phantom{22}$   & $  1.160  $ &  $        {-}0.890\phantom{1}$  & $0.30/7$ \\
$\pi^-$  	 &   $0.724\phantom{22}$   & $  1.160  $ &  $        {-}0.864\phantom{1}$  & $0.22/7$ \\
K$^+$    	 &   $0.0935\phantom{2}$   & $  1.187  $ &  $        {-} 0.174$   &  $0.07/7$   \\
K$^-$    	 &   $0.0927\phantom{2}$   & $  1.188  $ &  $        {-} 0.158$   &  $0.06/7$   \\
$\mathrm{p}$  &   $0.0432\phantom{2}$   & $  1.116  $ &  $        {-} 0.047$   &  $0.32/7$   \\
$\mathrm{\overline{p}}$  &   $0.0502\phantom{2}$   & $  1.089  $ &  $        {-} 0.088$   &  $0.26/7$   \\
$\Xi^-$    		&   $0.00552$   & $  1.108 $		&  $-0.043$              &  $0.12/1$   \\
$\Xibar^+$   	&   $0.00762$   & $  1.049 $		&  $-0.067$              &  $0.70/1$   \\
$\Omega^-$ 		&   $0.000286$             & $  1.324 $            &  $-0.0006$              &  $0.40/1$   \\
$\Omegabar^+$	&   $0.000309$             & $  1.308 $            &  $-0.0011$              &  $0.21/1$   \\
\hline
K$^{*0}/$K    &   \multicolumn{3}{c|}{see text for details}              &  $0.032/1$   \\
$\Lambda/\pi$  &   \multicolumn{3}{c|}{see text for details}              &  $0.0054/1$   \\ 

\end{tabular}
} 
\end{ruledtabular}
\end{table}

\subsection*{\label{ssec:ratios}Rebinning  K$^{0*}/$K$^-$,  $\Lambda/\pi$ hadron ratios}
We include particle ratios K$^{0*}/$K$^-$, $\phi/$K$^-$ and $\Lambda/\pi\equiv 2\Lambda/(\pi^-+\pi^+)$~\cite{Singha:2012qv}. This adds $\Lambda$, K$^{*0}$ and $\phi$ into our data set. Since in some ratios  certain systematic uncertainties of individual yields cancel out, introduction of ratios is reducing the overall error of the global fit. The ratio $\phi/$K has an experimental data point in each centrality bin we analyze, removing the need for interpolation or rebinning.
Thus the following only addresses K$^{0*}/$K$^-$, and $\Lambda/\pi$.

\begin{table}[!htb]
\caption{\label{tab:experimentalRatios} Experimentally measured ratios used as input to our interpolation as a function of $N_{\rm part}$.} 
\begin{ruledtabular}
\resizebox{\columnwidth}{!}{ 
\centering
\begin{tabular}{l|l|c|c|c}
 Ratio & Ref. & Centr. & $\langle N_{\rm part}\rangle$ & Experimental ratio   \\ \hline
\multirow{4}{1.5cm}{K$^{*0}/$K$^-$}&	\multirow{4}{*}{}   & 60--80\% & 22.6 & $0.333\pm 0.084$   \\
                               &   \cite{Singha:2012qv}                                           & 40--60\% & 68.8 & $0.285\pm 0.061$  \\
                               &                                              & 20--40\% & 157  & $0.245\pm 0.066$  \\
                               &                                              &  0--20\% & 308  & $0.194\pm 0.051$  \\
\hline
\multirow{4}{1.5cm}{$$\frac{2\Lambda}{(\pi^-\!\!+\pi^+)}$$}&	\multirow{4}{*}{}      & 60--80\% & 22.6  & $0.0355\pm 0.0041$  \\
                               &                                              & 40--60\% & 68.8 & $0.0371\pm 0.0042$  \\
                               &  \cite{Singha:2012qv}                                            & 20--40\% & 157  & $0.0365\pm 0.0042$  \\
                               &                                              & 10--20\% & 261  & $0.0355\pm 0.0041$  \\                               
                               &                                              &  0--10\% & 357  & $0.0336\pm 0.0040$  \\
\end{tabular}
} 
\end{ruledtabular}
\end{table}

There are four, resp. five, data points for K$^*/$K , resp. $\Lambda/\pi$, which we present in table~\ref{tab:experimentalRatios}. We describe K$^*/$K dependence on ${N_\mathrm{part}}$ with a power law
\begin{equation}
 \frac{{\rm K}^*}{{\rm K}} = f\left({N_\mathrm{part}}\right) = 16.23 ({N}_\mathrm{part}^{-0.0034} - 1) + 0.512,
 \label{eq:KstarOverK}
\end{equation}
with total $\chi^2/\mathrm{ndf} = 0.032/1$.
Systematic behavior of $\Lambda/\pi$ as a function of centrality is qualitatively different from K$^*/$K, see \rf{fig:KstarLambdaRatios}, a power law is not sufficient to properly describe the data. We use a sum of two power laws in the following form
\begin{align}
 \frac{\Lambda}{\pi} &= f\left({N_\mathrm{part}}\right)  \nonumber \\
 &= -6.79\times 10^{-5}\,{N}_\mathrm{part}^{0.848} -2\,{N}_\mathrm{part}^{-0.00135} + 2.03\,,
 \label{eq:LambdaOverPi}
\end{align}
with $\chi^2/\mathrm{ndf} = 0.0054/1$. In these two cases, the form of the ratio functions has no immediate physical meaning, it is invented in order to provide an accurate empirical description; note that  the bottom of the table~\ref{tab:interpolation} presents the fit quality of these two ratios.

Interpolation curves are depicted with dashed lines, in  \rf{fig:KstarLambdaRatios}, for K$^*$/K and $\Lambda/\pi$ ratios. To assign an error to the interpolated data points, we take the average nearby experimental error for the given particle yield or ratio. However, by extrapolating the K$^*/$K ratio to ${N_\mathrm{part}}=382$, we introduce systematic error due to our choice of the functional form of \req{eq:KstarOverK}.  To account for this effect, we multiply the error of K$^*/$K by $2$ (resp. $1.5$) in the most (resp. second to most) central bin we analyze as indicated by the shaded area in \rf{fig:KstarLambdaRatios}. As seen in \rf{fig:fittedparticles}, we also extrapolate $\Omega, \Xi$, but we do not believe that this adds to the already significant error, considering that  our power law interpolation functions describe other hadron yields up to $N_{\rm part}=382$.


\end{document}